\documentclass[11pt]{article}
\pdfoutput = 1
\usepackage[margin=1in]{geometry}
\usepackage{graphics, array}
\usepackage{multirow}
\usepackage{harvard}
\usepackage{pslatex}
\usepackage{paralist}
\usepackage{amsmath, amsthm, amssymb} 

\begin{document}
\title{\sc Consistent Valuation of Bespoke CDO Tranches}
\author{Yadong Li
\thanks{yadong.li@gmail.com: the views expressed in 
this paper are the author's own, and they do not necessarily reflect those 
of Barclays Capital. The author thanks Ariye Shater for many 
helpful comments.} \\
Barclays Capital}
\date{Mar 13, 2010}

\maketitle

\begin{abstract}
This paper describes a consistent and arbitrage-free pricing methodology 
for bespoke CDO tranches. The proposed method is a multi-factor extension to the 
\cite{self} model, and it is free of the known flaws in the current
standard pricing method of base correlation mapping. This method assigns a 
distinct market factor to each liquid credit index and models the correlation 
between these market factors explicitly.  A low-dimensional semi-analytical 
Monte Carlo is shown to be very efficient in computing the PVs and risks of 
bespoke tranches. Numerical examples show that resulting bespoke tranche 
prices are generally in line with the current standard method of 
base correlation with TLP mapping.  Practical issues such as model deltas 
and quanto adjustment are also discussed as numerical examples. 
\end{abstract}

\section{Introduction}
Base correlation mapping is the current market standard for pricing bespoke CDO tranches\footnote{
\cite{bcmap} is a detailed review of representative base correlation mapping methods.}.
Most market participants have implemented certain variations of the base 
correlation mapping method to price and risk manage bespoke tranches. Given the illiquid nature 
and wide bid/ask range of bespoke tranches, only rough agreements on the valuation of bespoke 
tranches between market participants are needed to facilitate trading. The standard base 
correlation mapping method played such a crucial role of providing the pricing consensus. 
The base correlation mapping method also produces 
stable risk measures, which allows a market participant to hedge bespoke tranches with 
index tranches and single name CDS contracts. The hedging 
activity of bespoke tranches in turn spurred further growth of the index tranche market. 
The base correlation mapping method therefore is a key enabler in the explosive 
growth of the bespoke and index tranche  markets in the years prior to the credit crisis.

When the credit crisis hits in 2008, the bespoke tranche market rapidly froze
because of the disappearing of the ``rating arbitrage'' and the lack of demand 
for illiquid and complex derivatives. However, the base correlation mapping method 
is still critical in practice for the risk management of the existing bespoke tranche 
positions. The losses from synthetic CDO tranche positions in major banks 
has been minuscule comparing to the write-downs from cash instruments such 
as ABS, MBS and CMBS bonds, which is a testimony of the effective risk management 
practice enabled by the base correlation model when comparing to the risk 
management practice in cash instruments.

Despite its popularity and practical importance, the base correlation mapping method 
is known to have serious flaws. The main problems include the 
lack of pricing consistency, existence of arbitrage and counter-intuitive risk measures. Some
of the flaws are well documented in \cite{bcdelta}.
The pricing of bespoke tranches are still of great practical interests because 1)
there are a large number of legacy bespoke tranche positions that require
risk management; 2) the standard index tranches remain liquid, and a bespoke 
tranche model is required for trading the standard index tranches since 
the index tranches eventually lose liquidity and behave like 
bespoke tranches as time goes by; 3) the consistent pricing of bespoke tranches 
is also important for developing relative value trading strategies between index
tranches from different indices and series. 

In this paper, we propose a consistent pricing method for bespoke CDO tranches 
that is free of the flaws in the base correlation model. This paper is organized 
as follows, we first review the base correlation mapping method; 
then we introduce the consistent pricing method for bespoke tranches based on 
the \cite{self} model; then we present the numerical results with special 
focuses on important practical issues such as tranche and single name deltas. 

\section{Review of Base Correlation Mapping\label{bcm}}
It is well known that the base correlation model cannot produce a consistent
joint distribution of default time or default indicators (abbreviated as
 $JDDI(t)$ and $\{JDDI(t)\}$ respectively following the notation in \cite{self}); 
therefore its practical application is limited to the computation of portfolio 
loss distributions. The lack of consistent $JDDT$ and $\{JDDI(t)\}$ in
the base correlation model has been frequently criticized by both 
practitioners and academics; however the base correlation model remains the 
market standard because it is a simple and flexible method to 
construct the portfolio loss distribution, which is by far the most 
important metric in practice.

The base correlation model is a very reliable interpolation method to price 
non-standard tranches on a liquid index portfolio, for example, an IG9 5-7\% 
tranche at 4Y maturity. Even though arbitrage could arise from the base 
correlation interpolation and extrapolation, the arbitrages on non-standard
tranches on liquid index portfolios are usually very small 
in practice since the base correlation interpolation and extrapolation 
methods have been exhaustively tweaked to minimize the arbitrage. The remaining
small arbitrage across index tranche strikes and maturities are usually not 
exploitable because of the wide bid/ask of the non-standard index tranches. 
Therefore, if we calibrate a base correlation 
model to standard index tranches, then use it to price tranches with non-standard 
strikes or maturities on the same index, the resulting prices are 
very consistent and reliable. Different dealers usually reach very 
similar non-standard index tranche prices even if their implementations 
of base correlation models differ in model assumptions and 
parameters\footnote{For example, 
different banks are likely to use different stochastic recovery 
specifications, and base correlation interpolation and extrapolation 
schemes}. 

In contrast, the bespoke tranches are priced with the more ad hoc base 
correlation mapping procedure, which leads to inconsistent loss distributions 
across different bespoke portfolios. This inconsistency is very
difficult to control and mitigate within the base correlation framework. 
The bespoke tranche pricing from base correlation mapping methods are
therefore much less reliable than the non-standard index tranches, and 
the risk measures computed from the base correlation mapping methods 
can be very misleading and counter-intuitive.

Here we briefly review the base correlation mapping methodology to
setup further discussions. We assume that a base correlation surface 
$\rho^I(t, K^I)$ has been calibrated to index tranches, where $K^I$
is the standard index strike. The superscripts $I, B$ are used to denote
index portfolio and the bespoke portfolio. The correlation surface of a 
bespoke portfolio $\rho^B(t, K^B)$ can be found via a mapping function $F$:
\begin{equation}
\label{mapping}
F(\rho, K^I, P^I(t), t) = F(\rho, K^B, P^B(t), t)
\end{equation}
Where $P^I(t), P^B(t)$ represents the index and bespoke portfolios,
including the notional amounts, default probabilities and recovery 
rates of all underlying names. Note that $\rho$ are the same on both 
side. Since $\rho$ is determined by $\rho = \rho^I(t, K^I)$ and 
$P^I(t), P^B(t)$ are all known, \eqref{mapping} defines an implicit 
function between $K^I$ and $K^B$ at time $t$. The mapping function $F$ is 
usually monotonic in both $K$ and $\rho$ to ensure the uniqueness
of the mapping between $K^I$ and $K^B$. A bespoke tranche is then 
priced with the base correlation of the mapped index strike: 
$\rho^B(t, K^B) = \rho^I(t, K^I)$. At-the-money (ATM), tranche loss 
proportion (TLP) and probability match (PM) are popular choices for the 
mapping function $F$, readers are referred to \cite{bcmap} for more 
details of these mapping functions. 

Currently there are multiple credit indices representing different regions, such
as CDX for the US and iTraxx for Europe. If a bespoke portfolio contains names 
from multiple geographical regions, the typical practice is to apply mapping 
to the credit indices of different regions separately, then the final price 
(or correlation) of the bespoke tranche is computed as the notional or risk 
weighted average of the prices (or correlations) of the different credit 
indices. 

The main problem of this base correlation mapping procedure is the
lack of identity consistency across portfolios. By ``identity 
consistency'', we mean that an issuer's default behavior should be 
the same no matter which portfolio it appears in; it should not change by 
merely moving from one portfolio to another portfolio. When computing
the loss distribution of the bespoke portfolio, the base correlation 
mapping method clearly violates the identity consistency because the mapping function 
$F$ depends on all the names in the portfolio. The ``weighting'' 
procedure to credit indices of multiple regions also violates the 
identity consistency because the weighting factors also depend on 
all other names in the bespoke portfolio.  The loss contribution from 
a unit notional of a specific name is therefore different across 
different portfolios because of the differences in the mapped 
bespoke correlation surfaces and the index weighting factors. 
The lack of identity consistency manifests itself as a number of 
practical problems. 

The first common problem from the lack of identity consistency is the 
arise of negative single name deltas. Intuitively a buy-protection position 
on tranche should increase in value if an underlying single name 
spread widens; however the change in correlation caused by the base 
correlation mapping could result in larger changes in the tranche's value 
than the change in spread, resulting in decrease in value of 
the buy-protection position. Negative single name delta is a serious 
challenge to risk management, as it leads to the absurd situation of 
needing to buy more protection on single names in order to hedge a 
buy-protection bespoke tranche position.

The dilemma of pricing fixed recovery tranche is another common 
problem. The protection payout in a fixed recovery tranche are 
computed using the recovery rate specified in the trade contract instead
of the market recovery rate determined by the dealer auction process 
after the name's default event. It has become a common practice to 
model the market recovery rate as stochastic; whereas the fixed 
tranche has to be modeled as deterministic recovery by definition. 
There can be (at least) two possible ways to price the fixed recovery 
tranche under base correlation mapping:
\begin{compactenum}
\item  apply the mapping with the stochastic market recovery, then 
use the resulting correlation surface to price the fixed recovery tranche.
\item apply the mapping with the fixed recovery 
rate, the mapping function thus uses the stochastic market recovery 
for the index portfolio and the fixed recovery for the bespoke portfolio. 
\end{compactenum}
The rationale of the first method is to price the stochastic market 
recovery tranche consistently with the fixed recovery tranche 
on the same bespoke portfolio. On the other hand, the fixed recovery 
portfolio could differ significantly in overall riskiness and 
dispersion from the same portfolio with stochastic market recovery. 
The rationale of the second method is to price the fixed 
recovery tranches consistently with other stochastic market recovery
tranches whose underlying portfolios have similar riskiness and dispersion
as the fixed recovery portfolio.  Both of these methods seem plausible but 
they give different prices. The cause of this dilemma is that the 
identity consistency is important when comparing the market recovery 
tranche with the fixed recovery tranche. Recognizing this, method 
1 attempts to preserve the identity consistency by using the same 
correlation between the market and fixed recovery bespoke tranche, 
which results in incompatible prices from the direct application of 
the base correlation mapping as in method 2, which only considers
the portfolio riskiness and dispersion, but not the identity
consistency. 

\section{Methodology}

The flaws in the base correlation mapping method described in the previous section
have been well known among researchers and practitioners. A number of
attempts have been made to develop better models to address the shortcomings
of the base correlation model. As pointed out in
\cite{self}, the minimum requirement for a model to price bespoke tranches 
consistently is to produce the joint distribution of default 
indicators ($\{JDDI(t)\}$). Traditional default time copulas, such as random 
factor loading in \cite{rfl} and the implied copula in 
\cite{impliedcopula} and \cite{skarke}, produce the joint 
distributions of default time ($JDDT$) whose marginal distribution is 
the $\{JDDI(t)\}$\footnote{Readers are referred to \cite{self} for more discussion 
on the $JDDT$ and $\{JDDI(t)\}$.}. Even though traditional default time copulas 
can be used for bespoke tranche pricing, they are very difficult 
to calibrate across multiple maturities because the global nature 
of the $JDDT$ induces strong coupling between different maturities. 
\cite{self} proposed the default indicator copula that
models the $\{JDDI(t)\}$ instead of $JDDT$, the advantage of default indicator
copula is that it is loosely coupled in time therefore it is much
easier to calibrate across multiple maturities. \cite{self} has shown that 
the default indicator copula can be easily calibrated to index tranches across
multiple maturities even near the peak of the credit crisis. In this
section, we propose a consistent method to price bespoke tranches
based on the default indicator copulas. The same 
methodology can be applied to default time copulas even though 
the default indicator copula and the $\{JDDI(t)\}$ are sufficient
for bespoke CDO pricing.

Once we have calibrated a default indicator copula to index tranches, 
it is a very natural step to take the 
calibrated copula and apply it to a bespoke portfolio. If there were 
only one liquid credit index in the market, the resulting bespoke pricing 
from this simple procedure would have been arbitrage free 
and had the desired property of identity consistency. However in reality, 
there are multiple credit indices 
whose tranches are traded liquidly, for example, CDX-IG, CDX-HY, and 
iTraxx are three main indices that each contains useful market information. 
A mixed bespoke portfolio that contains both investment grade and 
high yield names from the US and Europe should be priced using the market 
information from all three indices. The default indicator (or
default time) copulas are typically one factor for numerical efficiency, 
which is adequate for a single well-diversified credit index, but it is 
clearly inadequate in dealing with bespoke portfolios that contain 
names of different geographical regions and credit qualities. The 
one factor default indicator copulas have to be 
extended to multiple factors in order to properly price bespoke 
tranches. 

Recently there are several papers on the multi-factor approach
to price bespoke tranches. \cite{mcbspk} proposed a multi-factor 
default time copula that can be calibrated via a weighted Monte Carlo 
simulation method; \cite{halpbspk} proposed a hybrid multi-factor
model that mixes the top-down and bottom-up approach. Both of these 
methods assumes there are multiple common market factors behind
each credit index. In \cite{mcbspk}, the factors are 
economic factors, such as geographical region, industry and sector. In 
\cite{halpbspk}, the factors are associated with whether a name 
appear in the bespoke tranche we want to price. Under these
multi-factor extensions, the calibration to index tranches becomes
more time-consuming because of the additional factors. \cite{mcbspk}
used a weighted Monte Carlo simulation to calibrate the joint
distribution of all factors; \cite{halpbspk} used a two step approach
that a bottom up model is used to compute the prior loss 
distribution first; then a top-down method is used to adjust the prior
distribution to match the observed index tranche prices. Both of
these methods are still subjected to certain practical limitations: 
for example, \cite{mcbspk} only showed calibration and pricing results 
for a single maturity; it is difficult numerically to extend
the same Monte-Carlo based calibration method to multiple maturities
because of the strong coupling of the default time copula across
different maturities. The \cite{halpbspk} approach, on the other hand, 
is not fully bottom-up, so the identity consistency is not fully 
preserved, and it can be cumbersome in dealing with real ``bespoke'' 
names that do not belong to any credit index. 

\subsection{Multi-factor Extension}
In this section, a very simple and efficient multi-factor extension
to the \cite{self} model is described. This method is fully consistent 
and arbitrage free across both time and capital structure; and it
preserves identity consistency by construction. This pricing method 
is free of all the flaws in the standard base correlation mapping 
methods, for example: its single name deltas are always positive 
and there is no ambiguity in the pricing of fixed recovery tranches.

Even though it is very important for a model to produce arbitrage free 
prices, there are other requirements that are also
important in practice, for example: the model has to produce reasonable 
index tranche deltas for market making purposes, and the model's 
bespoke and off-the-run index tranche prices should not be too far away
from the current market consensus as determined by the base correlation
model with the TLP mapping (abbreviated as BC-TLP). Even if a model can 
produce intrinsically consistent and arbitrage free prices, its acceptance
would be difficult if its prices and tranche deltas are too far away from 
the current market observations and consensus. Also, the model has to be
fast enough to compute the PVs and risks of a large number of 
bespoke tranches in a reasonable time before it can be used in practice.
Later in this article, we will show that the proposed method does meet 
these practical requirements.

Readers are assumed to be familiar with the modelling framework
described in \cite{self}, where it showed how a one-factor default
indicator copula can be calibrated to index tranches 
across multiple maturities. Unlike the \cite{mcbspk} and \cite{halpbspk}
where each credit index can be affected by multiple market factors,
here we make the assumption that there is a single distinct market factor 
for each credit index. The reason behind this assumption is that the index 
portfolios such as CDX and iTraxx are well diversified therefore 
its tranche prices only reflect the broad market factor rather 
than any of the specific sector or industry factors. The same 
argument is also made in \cite{mcbspk}.
The advantage of this one-to-one mapping between market factors 
and liquid credit indices is that it does not add any complexity
to the calibration procedure to index tranches, each credit index
can be  calibrated separately. 
Itraxx-S9, CDX-IG9 and CDX-HY9 are the three most important liquid
credit indices in the current market\footnote{The most liquid series 
may not be the latest series, for example, CDX-IG9 is still the most 
liquid CDX series while the CDX-IG13 is the latest series.};
we need to consider at least these three market factors in order
to cover most bespoke tranches in practice. Because
of the one-to-one mapping between market factor and 
liquid indices, this approach is easily extensible if other credit 
indices become liquid. 

Given the globalization of world economy, the market factors behind
these different credit indices are expected to be highly correlated: if US
issuers are suffering large number of defaults we expect the same
in European issuers, and vise versa.  The correlation between 
different market factors (denoted as $X^i_t$ for the $i$-th market 
factor) is a key risk factor for the mixed bespoke portfolios 
containing names across multiple regions. There are many possible ways 
to generate correlated $X^i_t$ processes from their calibrated marginal
distributions. Since we only need the joint distribution of market 
factors at each time grid for the purpose of bespoke 
tranche pricing, we can take a very simple approach that models the 
correlation between market factors using a static copula function (e.g.,  the 
classic Gaussian Copula). Under the \cite{self} framework, the
marginal distribution of a single market factor $X^i_t$ is known 
from the calibration to its associated index tranches. Therefore, 
the joint distribution of the market factors is fully 
specified once the copula function is chosen. The time 
consistency is also preserved by construction as long as the same 
copula function is applied to all times.

Since the identities and the joint distributions of common market 
factors are fully specified by the liquid credit indices and the copula 
function, the bespoke tranche prices are fully determined if we can 
establish a fixed relationship between every issuer 
and every market factors. The fixed relationship between issuers and 
market factors ensures the identity consistency because this relationship
does not depend on any other names; therefore a given issuer 
behaves the same way no matter which portfolio it appears. Given
the one-to-one mapping between market factors and liquid indices,
issuers that are part of a liquid index can only be exposed to the
market factor of that index. But issuers that are not part of
any liquid indices can be exposed to multiple market factors in the
general case. For example, a conglomerate that operates in both the 
US and Europe may be sensitive to both of the CDX and iTraxx 
market factors. A multi-factor extension to the \cite{self} model
is therefore needed. The following is a multi-factor version of the 
conditional default probability function that can be used 
in Definition 3.4 of \cite{self}:
\begin{equation}
\label{condp}
p_j(\vec{X_t}, t) = 1 - c_j(t)e^{-b_j(t) \sum_i\beta^i_j X^i_t}
\end{equation}
where the subscript $j$ identifies the issuer and the superscript $i$ 
identifies the market factor. We use $\vec{X_t}$ to denote the vector of 
all the market factors. The $\beta^i_j$ are constant factors that controls the 
relative exposure weights to different market factors. The $b_j(t)$ is 
a time-dependent scaling factor so that the overall systemic contribution 
is a constant fraction of the unconditional cumulative hazard:
\begin{equation}
\label{sncal}
\log(\mathbb{E}[e^{-b_j(t)\sum_i\beta^i_j X^i_t}]) = \gamma_j(t) \log(1-p_j(t))
\end{equation}
where $p_j(t)$ is the unconditional default probability extracted from
single name CDS curve; the $-\log(1-p_j(t))$ is the cumulative hazard of 
the $j$-th name at time $t$; the $\gamma_j(t)$ is a time-varying fraction 
for the systemic factor contribution. The $c_j(t)$ is the idiosyncratic 
factor contribution which makes up the rest of the cumulative hazard:
\[
\log(c_j(t)) = (1-\gamma_j(t)) \log(1-p_j(t))
\]
Because the market factors are correlated, the expectation on the LHS 
of \eqref{sncal} has to be computed via a multi-dimensional integration, 
which may require a Monte Carlo simulation if the number of market factors are
more than two. Therefore, the calibration of $b_j(t)$ factor is more 
time-consuming than in the single factor case. However, the $b_j(t)$
only needs to be computed once for each issuer, we can improve the computational
efficiency by pre-computing and caching all the $b_j(t)$ factors before valuing 
all the bespoke tranches. Under this multi-factor extension, the default 
indicators are independent conditioned on the vector of $\vec{X_t}$. 

The specification in \eqref{condp} has a simple econometric explanation
if we re-write it in a different form:
\begin{equation}
\label{mfreg}
- \log(1-p_j(\vec{X_t}, t)) = -b_j(t) \sum_i\beta^i_j X^i_t + \log(c_j(t))
\end{equation}
which is simply a multi-factor linear regression of a name's 
cumulative hazard (the LHS) using market factors as explanatory
variables, the $b_j(t)\beta_j^i$ is the factor loading to the market
factor $X_t^i$ and the $\log(c_j(t))$ is the residual idiosyncratic error 
that is specific to the j-th name.

Once we established the fixed relationship between all the issuers and
market factors (i.e.: all the $\beta_j^i$s and $\beta_j(t)$s), 
all the bespoke tranche prices are uniquely determined. This method of 
pricing bespoke tranches is fully bottom-up, and it does not involve 
any ad hoc mapping method. With this pricing method, the only
sources of uncertainty in the bespoke tranche prices are the relative 
factor loadings of $\beta_j^i$, the systemic fraction $\gamma_j$, and
the correlation between market factors. All of these parameters have
clear meanings and are closely related to market observables, therefore 
they can be easily estimate from the historical data. For example, the 
correlation between market factors can be estimated from historical 
spread movements of their corresponding indices; the relative factor 
loading $\beta_j^i$ can be estimated by regressing the single name's
spread movement to spread movements of relevant credit indices.

\subsection{Semi-analytical Monte Carlo\label{samc}}
A straight-forward Monte Carlo simulation of default time can be used 
to price bespoke tranches under the multi-factor extension. The easiest
way to run a default time simulation is to use a co-monotonic\footnote{See
\cite{losslinker} for more details on co-monotonic Markov Chain.}
Markov Chain on the common market factors:
\begin{center}
\underline{Monte Carlo Simulation of Default Time}
\end{center}
\begin{compactenum}
\item Draw a set of correlated uniform random numbers $\vec{u}$ from the
copula function of market factors; note that the $\vec{u}$ has the same
dimensionality as $\vec{X_t}$.
\item At each time horizon $t$, invert each element of the correlated 
$\vec{u}$ to the corresponding market factor $X_t^i$ using the marginal 
distributions of the market factor. Because the same $\vec{u}$ is used for 
all time horizon, the resulting market factor paths over time are 
effectively co-monotonic. Note that the market factor paths drawn this 
way are always monotonically increasing, therefore, the time consistency 
is fully preserved.
\item Draw the individual name's default time from the conditional 
default probabilities in \eqref{condp} separately as the default times
are independent conditioned on the path of $\vec{X_t}$ over time. 
Also draw the recovery rate if stochastic recovery model is used.
\item Compute the tranche payoff from the simulated default times, and 
average over many paths to get the tranche PV. 
\end{compactenum}
The modelling framework of \cite{self} guarantees that we would
obtained the same tranche prices using other Markov 
chains since the tranche price only depends on the $\{JDDI(t)\}$,
which is not affected by the choice of the Markov chain.

This default time simulation procedure is of theoretical interest
since it evidenced that the suggested multi-factor extension is fully 
consistent and arbitrage free; however it 
is notoriously slow in computing single name risk measures because 
of the digital nature of the default event. The small perturbation in 
a single-name's default probability only causes meaningful change in 
the tranche payoff if it happened to move the default time across
the trade maturity. For vast majority of the simulated paths, 
small perturbation in a single name's default probability does not
cause any difference in the tranche payoff, therefore it requires a 
huge number of simulation paths for the single name risks to converge. 
Given the importance of single name risks, this default time simulation
is clearly inadequate for practical purposes. 

A better Monte Carlo simulation procedure can be devised by extending
the standard one-factor semi-analytical pricing technique as
described in \cite{semianalytical} to multiple factors. Since the default 
indicators are independent conditioned on $\vec{X}_t$,
the conditional tranche losses at time $t$ can be computed analytically  
(for example, using normal approximation). For the given time $t$, the 
unconditional ETL under the multi-factor model can therefore be obtained 
by integrating the conditional ETL over the joint distribution of market 
factors, which requires a low-dimensional Monte Carlo simulation. 
We call this method the Semi-analytical Monte Carlo method. The following 
is the detailed steps of computing the ETL at a given time horizon $t$ 
using the semi-analytical Monte Carlo simulation:
\begin{center}
\underline{Semi-Analytical Monte Carlo Simulation}
\end{center}
\begin{compactenum}
\item Draw a vector of $\vec{X_t}$ at the time $t$ according to the
copula function of the market factors. 
\item For each issuer, compute the conditional default probability 
from \eqref{condp}, and the conditional recovery rate if
stochastic recovery is used.
\item Compute the conditional mean and variance of portfolio
losses, and use normal approximation to compute the conditional
ETL, see \cite{self} for more details on the normal approximation.
\item Average the conditional ETL over many draws of $\vec{X_t}$
to obtained the unconditional ETL
\end{compactenum}
The above procedure can be repeated for each quarterly time grid
to construct the full ETL curve of the tranche, and the tranche price
immediately follows. This semi-analytical Monte Carlo simulation is a 
very natural extension from the typical one factor semi-analytical method. 
A similar simulation technique is also used in \cite{mcbspk}.

This semi-analytical Monte Carlo method is
much more efficient than the simple Monte Carlo simulation of default 
indicators since its dimensionality (which is the number of market factors) 
is much smaller than the dimensionality of the default time 
simulation (which is the number of names in the bespoke portfolio).  This 
semi-analytical Monte Carlo method can compute single name deltas 
very efficiently because the effects of single name perturbation 
are analytically captured in the conditional mean and variance of 
portfolio, which leads to a steady change in the conditional ETL 
for each simulated path.

\subsection{Many-to-One Restriction\label{many1}}
A even more efficient semi-analytical Monte Carlo method can be devised if 
we adopt the restriction that each issuer can only be associated 
with a single market factor, ie, the relationship between issuers
and market factors is many-to-one instead of many-to-many. The
many-to-one restriction is acceptable in practice because 1) most 
issuers' main operations are within one geographical region, 2) the 
market factors are highly correlated to each other therefore those 
issuers operating in multiple regions can be well approximated 
using the market factor of its main operation. The second argument 
is more apparent with the regression form of the conditional
default probability specification in \eqref{mfreg}: in a multi-factor
regression analysis, there is only a small impact to the fit quality 
if we ignore factors that are highly correlated to existing 
factors.

The following is a possible procedure to associate issuers to a 
single market factor:
\begin{compactenum}
\item Issuers that are part of a liquid index are only sensitive to 
the market factor of that index. 
\item Issuers whose main business is in Europe are only sensitive to 
the market factor of iTraxx. 
\item For issuers whose main business is in US, the investment grade 
names are only sensitivity to the CDX-IG9, and the high yield issuers 
are only sensitivity to CDX-HY9.
\end{compactenum} 
With the many-to-one restriction, the bespoke tranches can be priced 
extremely fast. First, we no longer need to run Monte Carlo simulation to 
calibrate the $b_j(t)$ in \eqref{sncal}. Secondly, we can greatly improve 
the efficiency of the semi-analytical Monte Carlo as the follows:
\begin{center}
\underline{Semi-Analytical Monte Carlo Simulation with Many-to-One Restriction}
\end{center}
\begin{compactenum}
\item Divide he bespoke portfolio into sub-portfolios according
to market factors. Each name can only appear in one sub-portfolio 
because of the many-to-one restriction. 
\item 
We use $L^i_t$ to denote the loss in the sub-portfolio associated with the 
$i-$th market factor. Since names in each sub-portfolio are exposed to the same 
market factor, the conditional mean and variance of each sub-portfolio's 
loss (denoted as $\mu_i(L^i_t | X^i_t)$ and $\sigma^2_i(L^i_t | X^i_t)$) can be pre-computed 
and cached for each possible value of $X_t^i$ (here we assume 
that the distribution of market factors $X_t^i$ can be discretely sampled).
\item The conditional loss distribution of the total portfolio can then be constructed 
from the conditional loss distribution of the sub-portfolios. Since the loss of the 
whole portfolio is the sum of the losses from sub-portfolio ($L_t = \sum_i L^i_t$) 
and the sub-portfolio losses are independent conditioned on $\vec{X_t}$, 
the conditional mean and variance of the whole portfolio loss are
simply the summation of those of the sub-portfolios: $\mu(L_t | \vec{X_t}) = \sum_i \mu_i(L^i_t | X^i_t)$,
$\sigma^2(L_t | \vec{X_t}) = \sum_i \sigma_i^2(L^i_t | X^i_t)$.
The conditional expected tranche loss (ETL) of the whole portfolio 
can then be obtained using normal approximation. 
\item The unconditional ETL can then be computed by integrating the 
conditional ETL over the multi-dimensional distribution of the market 
factors via a low-dimensional Monte Carlo simulation. 
\end{compactenum}
Since the conditional mean and variance are pre-computed and cached
for each sub-portfolio, the above Monte Carlo simulation is extremely
fast. The many-to-one restriction may seem restrictive for issuers 
that operate in multiple regions. However, given that the majority 
of issuers mainly operate in one geographical region, it is a good 
trade off in practice since this restriction leads to an extremely
simple and efficient method to price bespoke tranches, and it still
captures the main risk factors in bespoke tranches.

In the following discussions, we use the abbreviation DIC and SAMC to
refer to the default indicator copula and the semi-analytical Monte
Carlo pricing method for bespoke tranches.

\section{Numerical Results}
In this section, we present some numerical results of bespoke tranche
pricing with DIC-SAMC, with special attention to practical issues such 
as deltas and comparisons to the BC-TLP model.

The market data is taken from Dec. 31st, 2009. The DIC is calibrated 
to the expected tranche loss of CDX-IG9, iTraxx-S9 and CDX-HY9 separately. 
The details of the calibration procedure is described in \cite{self}. 
Since the standard maturities of iTraxx-S9 are half year longer 
than those of CDX-IG9 and CDX-HY9, for the purpose of illustrating 
the bespoke pricing with mixed portfolios, we calibrated the iTraxx-S9 
tranches at the standard maturities of CDX-IG9 and CDX-HY9. 
Since the ETLs are not directly observable, a standard base correlation 
model is used to extract the ETLs from index tranche prices; the
ETLs are then used as calibration input to the DIC model. For the rest 
of this paper, all tenors refer to those of the 
CDX-IG9 index, e.g. 5Y means Dec 20, 2012, and 7Y means Dec 20, 2014.

\subsection{$\gamma_j(t)$ and Model Calibration}
\begin{figure}
\caption{The Systemic Fraction - $\gamma_j(t)$\label{sf}}
\center
\scalebox{.65}{\includegraphics{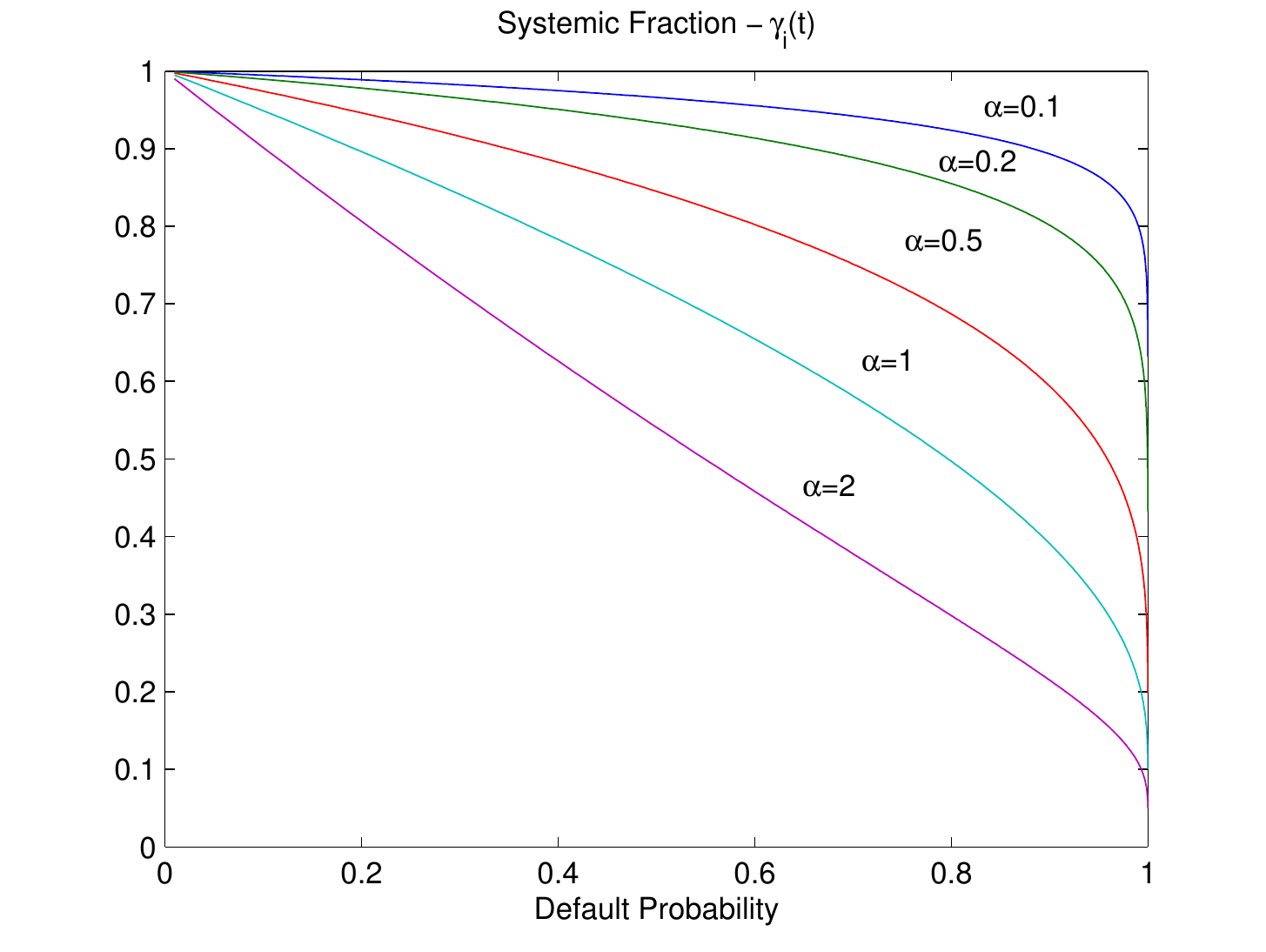}}
\end{figure}

Before we discuss the bespoke tranche pricing, we need to choose
the $\gamma_i(t)$ in \eqref{condp} which is the fraction
of systemic factors' contribution to the overall cumulative hazard. 
This parameter is of importance because it affects the 
calibration and resulting bespoke prices. Instead of estimating 
$\gamma_j(t)$ from historical time series, we choose a very simple 
functional form for the $\gamma_j(t)$ parameter. It is a common 
belief that riskier names are more sensitive to idiosyncratic risk 
and safer names are more sensitive to the systemic risk; therefore 
we choose the marginal systemic fraction to be a decreasing 
function of the cumulative hazard. We denote the cumulative hazard
as $h_j = -\log(1-p_j(t))$ for the $j$-th name. A simple exponential 
function $e^{-\alpha h_j}$ is convenient for the marginal systemic 
fraction since it is always within 0 and 1; and the $\alpha$ is a free parameter 
that controls how fast the marginal systemic faction decrease 
with the cumulative hazard. Then $\gamma_i(t)$, being the cumulative 
systemic fraction, can be computed from the marginal systemic fraction: 
\begin{equation}
\label{gamma}
\gamma_j(t) = \frac{1}{h_j} \int_0^{h_j} e^{-\alpha h} d h
 = \frac{1-e^{-\alpha h_j}}{\alpha h_j}
\end{equation}
For simplicity, we choose to use the same $\alpha$ parameter 
for all issuers even though they could be name specific. Figure 
\ref{sf} showed the $\gamma_j(t)$ for different $\alpha$ values.

\begin{figure}
\caption{DIC Calibration - $X_t$ CDF for CDX-IG9\label{xcal}}
\vspace{.5cm}

\center
\begin{minipage}{3in}
\center
\underline{$\alpha=0.2$}
\scalebox{.55}{\includegraphics{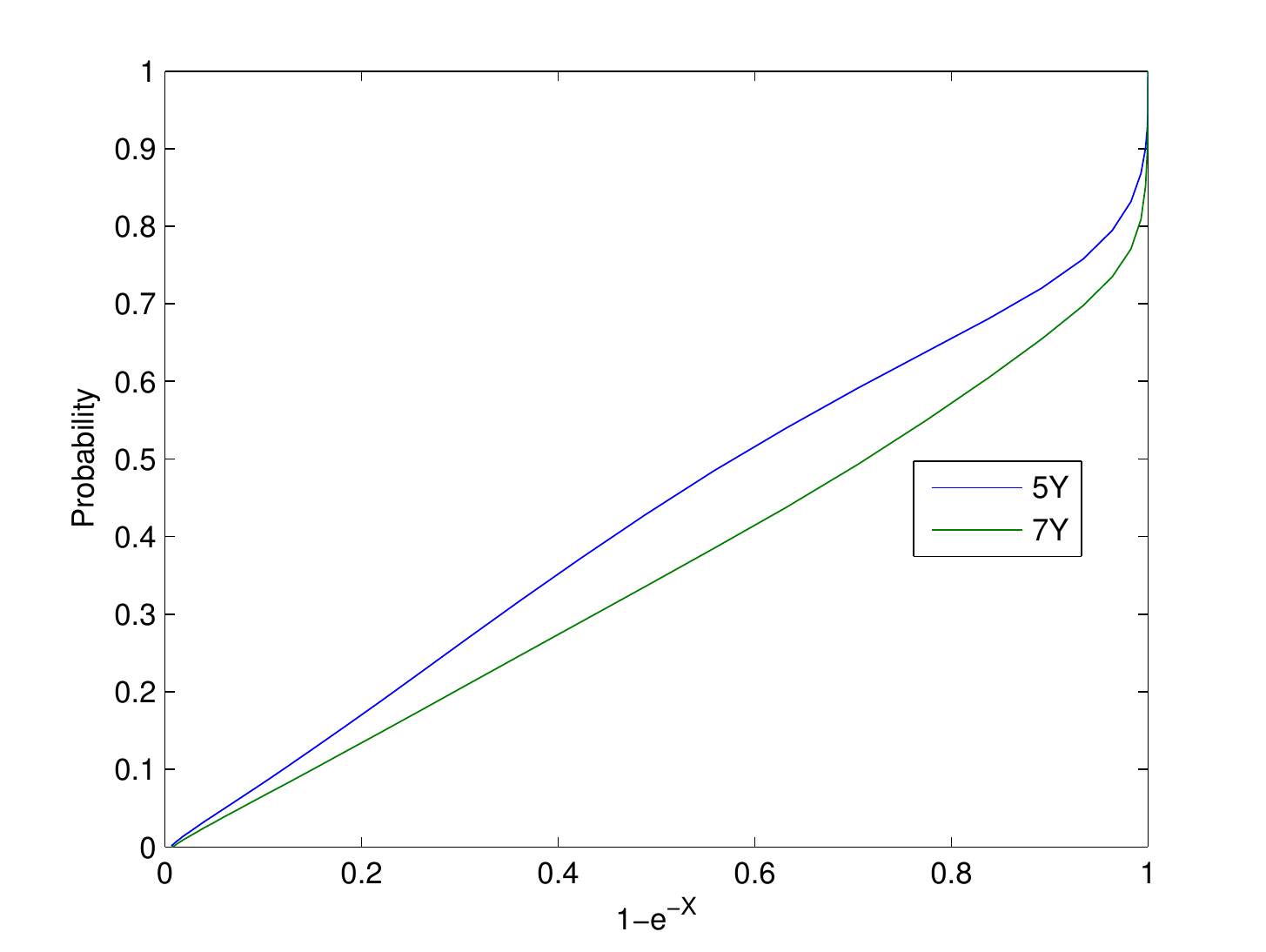}}
\end{minipage}
\begin{minipage}{3in}
\center
\underline{$\alpha=1$}
\scalebox{.55}{\includegraphics{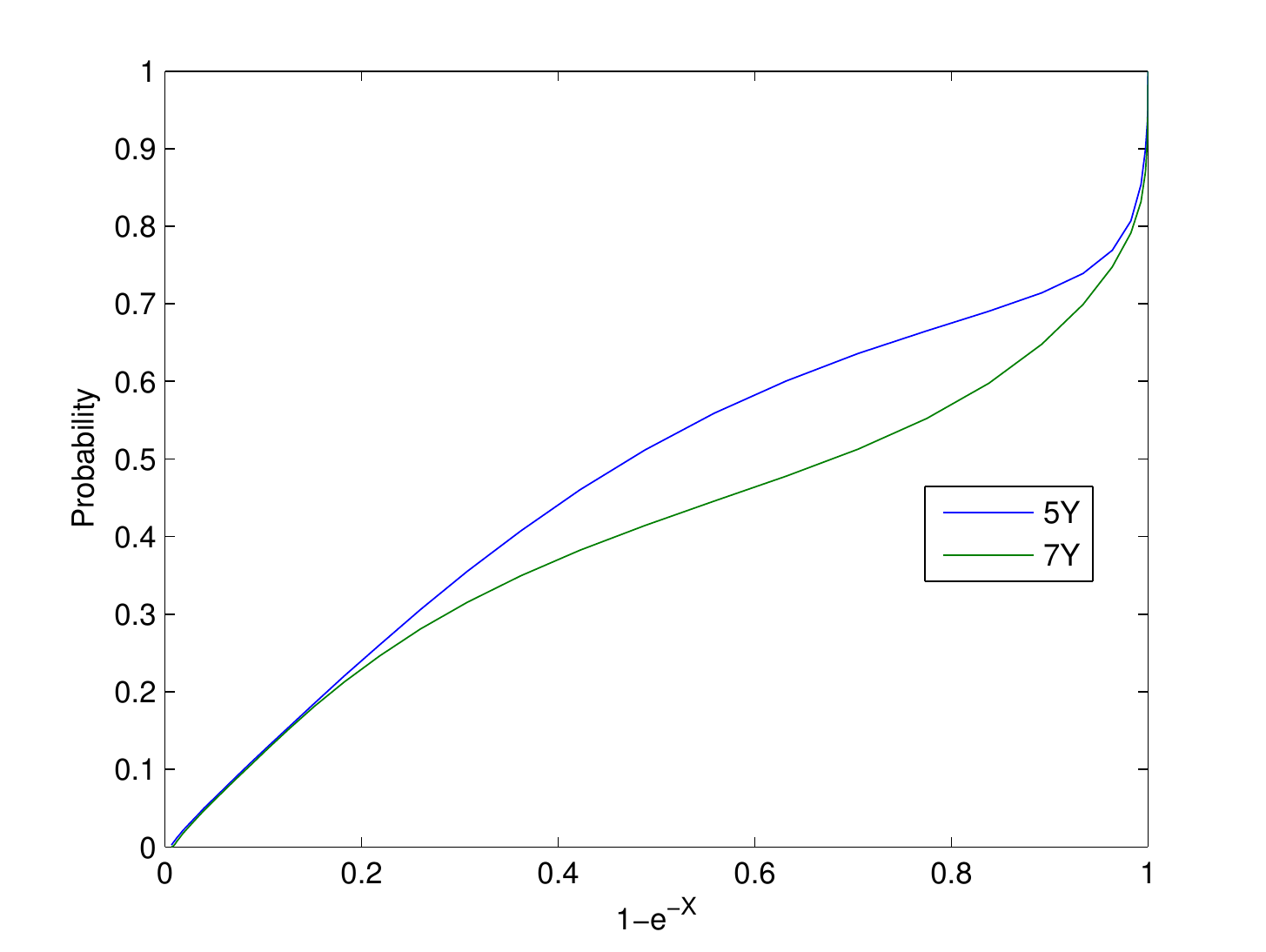}}
\end{minipage}
\end{figure}

The DIC is calibrated to both the 5Y and 7Y ETL of CDX-IG9, CDX-HY9 
and iTraxx-S9 with two different values of $\alpha = 0.2$ and 
$\alpha = 1$. Only the 5Y and 7Y tenors are used in this study 
because all three indices trade liquidly at these two 
tenors. The model can be calibrated very closely to 
input ETLs of all three indices with either $\alpha$ value, 
table \ref{licalib} showed the ETL input and the DIC model fit. Figure
\ref{xcal} showed the calibrated distribution of the common factor
$X_t$ for the CDX-IG9 index. The calibrated distributions of $X_t$
are well behaved for both $\alpha$ values. The calibrated
CDF curves at the two maturities never cross each other because 
the common factor $X_t$ in the \cite{self} framework has to be an
increasing process, and this monotonic constraint is built into
the calibration procedure.

\begin{table}
\caption{DIC Calibration - Model Fit on Dec 31, 2009 \label{licalib}}

\center
\footnotesize

\begin{tabular}{|rr|rr|rr|rr|}
\hline
\multicolumn{ 2}{|c|}{{\bf CDX-IG9}} & \multicolumn{ 2}{|c|}{{\bf Market ETL}} & \multicolumn{ 2}{|c|}{{\bf Fit with $\alpha=0.2$}} & \multicolumn{ 2}{|c|}{{\bf Fit with $\alpha=1$}} \\
\hline
 {\bf Att} &  {\bf Det} &   {\bf 5Y} &   {\bf 7Y} &   {\bf 5Y} &   {\bf 7Y} &   {\bf 5Y} &   {\bf 7Y} \\
\hline
     0.0\% &      2.4\% &    67.15\% &    82.14\% &    67.23\% &    82.18\% &    67.22\% &    82.41\% \\

     2.4\% &      6.5\% &    23.20\% &    42.42\% &    23.24\% &    42.40\% &    23.24\% &    42.56\% \\

     6.5\% &      9.6\% &     7.52\% &    21.79\% &     7.53\% &    21.71\% &     7.53\% &    21.86\% \\

     9.6\% &     14.8\% &     3.17\% &    10.30\% &     3.18\% &    10.19\% &     3.18\% &    10.24\% \\

    14.8\% &     30.3\% &     0.81\% &     2.68\% &     0.83\% &     2.80\% &     0.83\% &     2.92\% \\

    30.3\% &     61.2\% &     0.23\% &     1.90\% &     0.33\% &     1.86\% &     0.33\% &     1.77\% \\
\hline
\end{tabular}  

\vspace{.5cm}

\begin{tabular}{|rr|rr|rr|rr|}
\hline
\multicolumn{ 2}{|c|}{{\bf iTraxx-S9}} & \multicolumn{ 2}{|c|}{{\bf Market ETL}} & \multicolumn{ 2}{|c|}{{\bf Fit with $\alpha=0.2$}} & \multicolumn{ 2}{|c|}{{\bf Fit with $\alpha=1$}} \\
\hline
 {\bf Att} &  {\bf Det} &   {\bf 5Y} &   {\bf 7Y} &   {\bf 5Y} &   {\bf 7Y} &   {\bf 5Y} &   {\bf 7Y} \\
\hline
     0.0\% &      3.0\% &    35.59\% &    56.69\% &    35.60\% &    55.72\% &    35.60\% &    56.77\% \\

     3.0\% &      6.0\% &     7.73\% &    22.06\% &     7.73\% &    21.70\% &     7.73\% &    22.19\% \\

     6.0\% &      9.0\% &     4.56\% &    13.46\% &     4.56\% &    13.12\% &     4.56\% &    13.55\% \\

     9.0\% &     12.0\% &     2.27\% &     7.31\% &     2.27\% &     7.05\% &     2.27\% &     7.39\% \\

    12.0\% &     22.0\% &     0.80\% &     3.02\% &     0.80\% &     3.06\% &     0.80\% &     3.01\% \\

    22.0\% &     60.0\% &     0.51\% &     1.84\% &     0.53\% &     1.97\% &     0.53\% &     1.84\% \\
\hline
\end{tabular}  

\vspace{.25cm}

\begin{tabular}{|rr|rr|rr|rr|}
\hline
\multicolumn{ 2}{|c|}{{\bf CDX-HY9}} & \multicolumn{ 2}{|c|}{{\bf Market ETL}} & \multicolumn{ 2}{|c|}{{\bf Fit with $\alpha=0.2$}} & \multicolumn{ 2}{|c|}{{\bf Fit with $\alpha=1$}} \\
\hline
 {\bf Att} &  {\bf Det} &   {\bf 5Y} &   {\bf 7Y} &   {\bf 5Y} &   {\bf 7Y} &   {\bf 5Y} &   {\bf 7Y} \\
\hline
     0.0\% &      4.0\% &    78.29\% &    91.76\% &    78.29\% &    91.67\% &    78.29\% &    92.15\% \\

     4.0\% &     15.6\% &    36.47\% &    62.78\% &    36.47\% &    62.54\% &    36.48\% &    62.54\% \\

    15.6\% &     27.2\% &    13.20\% &    32.05\% &    13.20\% &    31.81\% &    13.20\% &    32.00\% \\

    27.2\% &     56.3\% &     4.55\% &    21.63\% &     4.55\% &    21.42\% &     4.55\% &    21.49\% \\
\hline
\end{tabular}  
\end{table}

\begin{figure}
\caption{Calibrated Base Correlation Curve\label{bccal}}
\vspace{.25cm}

\center
\begin{minipage}{3in}
\center
\underline{5Y}
\scalebox{.55}{\includegraphics{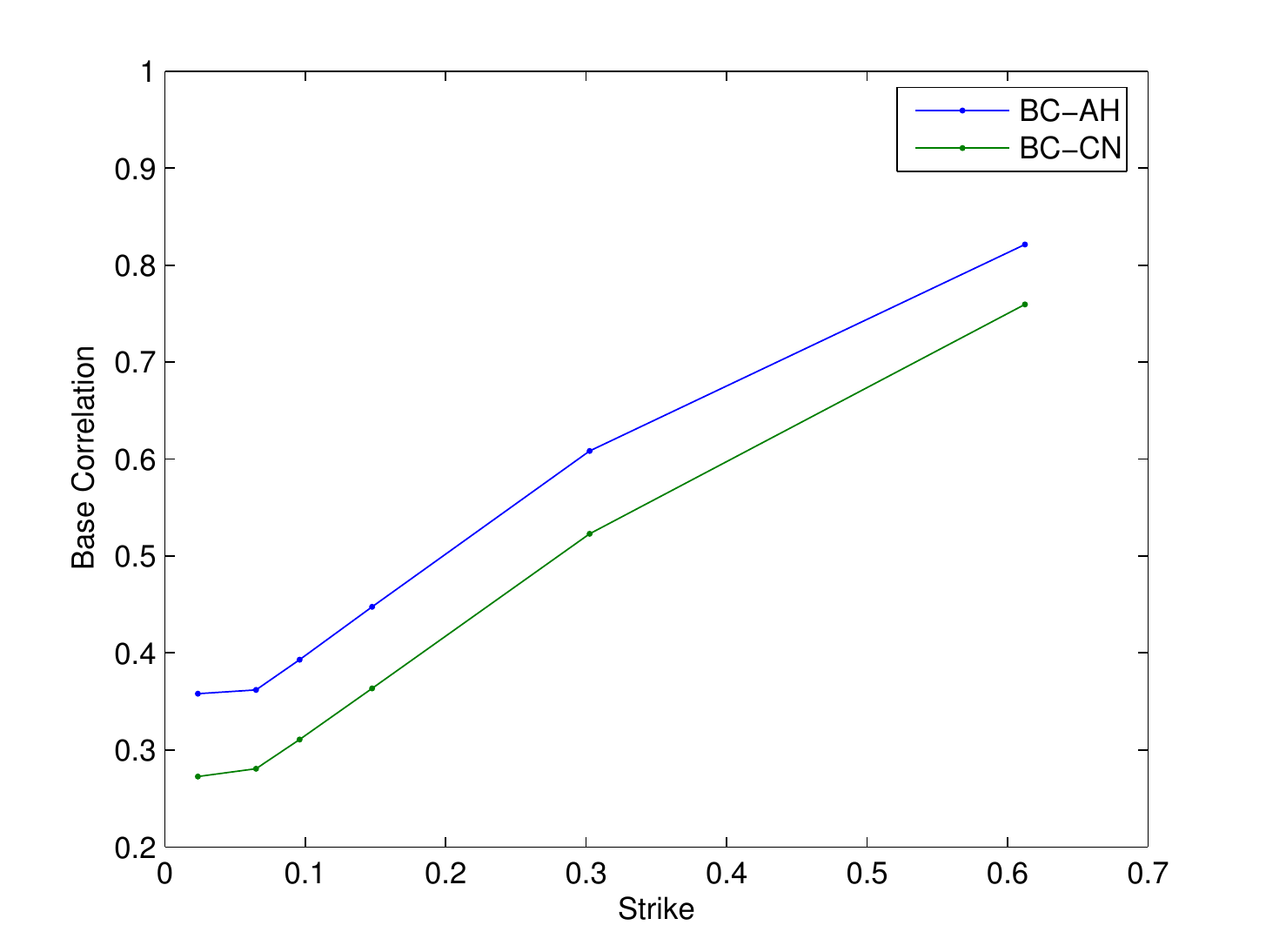}}
\end{minipage}
\begin{minipage}{3in}
\center
\underline{7Y}
\scalebox{.55}{\includegraphics{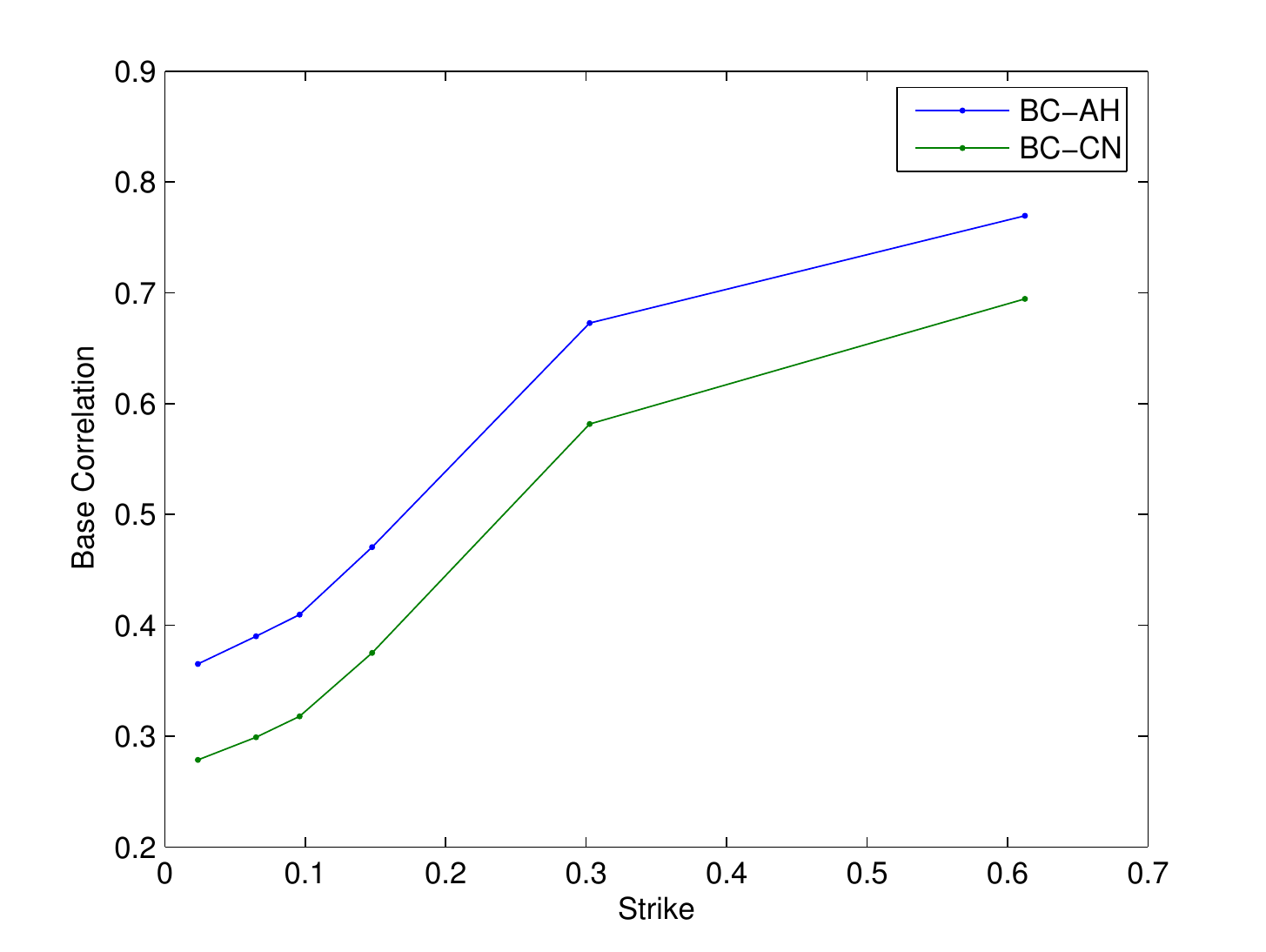}}
\end{minipage}
\end{figure}

For comparison purposes, the base correlation model is also calibrated 
to the three indices. Two different stochastic recovery specifications
are used, the first is the \cite{hitier} method with a recovery markdown
factor of 0.01, the second is the cumulative normal recovery proposed in
the \cite{rfl}. With different stochastic recovery 
specifications, the calibrated base correlation curves are different,
as shown in Figure \ref{bccal}, where BC-AH and BC-CN represents the
two different stochastic recovery models.

\subsection{Index Tranche Delta}
A practical concern of moving away from the market standard BC-TLP 
model is the changes in model deltas. 
We first give a brief description on the index tranche delta for
readers who are not familiar with the index tranche market. The index 
tranches are commonly traded in a delta-hedged package, i.e., when
a client enters a tranche trade with a dealer, he also trades 
a hedging index swap with the same dealer; so the dealer's
book is delta hedged with the new trade. The hedge ratio is 
published by the dealer as part of the index tranche quote, which is 
normally referred as the ``tranche delta'' or ``leverage ratio''. The 
index swaps are much more liquid than the index tranches, and their 
prices are updated much more frequently throughout a trading day. With this 
delta-exchange mechanism, an index tranche trader does not need to 
publish new index tranche quotes every time when the index swap 
spread moves; he only need to do so if his current level 
is different from the delta-adjusted prices of the previous quote. 
This delta exchange mechanism implies that the index tranche prices 
are the delta-adjusted levels for small movements in the 
index swap spreads.

We use ``market delta'' to refer to the index tranche deltas quoted 
by the market makers and ``model delta'' to refer to the tranche delta calculated
by a model. The model delta is usually computed by bumping the index swap spreads, 
re-applying the swap adjustment and keeping the model's correlation parameters 
constant\footnote{The market index swap spread is usually different from 
the ``intrinsic'' spread computed from the the underlying single name CDS 
spreads. This basis between intrinsic and market index swap spread can be
quite large when market is in distress. Before index tranche pricing and 
calibration, the underlying single name spreads are usually adjusted to match 
the current market index swap spread because the market index swap spread is more 
liquid than the underlying single names' CDS spreads. This procedure is 
commonly referred as ``index swap adjustment''.}.  \cite{bardelta} 
has shown that the model delta is dependent upon how the single name 
spreads are bumped during the index swap adjustment. For example, an additive 
bump where the bump sizes are constant across all names produces very
different model deltas from a multiplicative bump where the bump sizes are 
proportional to the current spread levels. Currently the CDX and iTraxx market 
deltas are closer to the base correlation model deltas with additive bumps than 
the multiplicative bumps.

The market tranche deltas are typically determined by a dealer poll process: 
each major dealer submits their market delta quotes to a dealer poll at the
market open, and the averages of all dealers' submission are then used by all 
the dealers as the market deltas for that day. Individual dealer usually submit 
their model delta to the dealer poll, which can be quite different from each 
other due to difference in model and swap adjustment assumptions. This dealer 
poll process helps improve the transparency in the index tranche market by 
ensuring that all major dealers are trading with similar tranche deltas despite 
their model differences. Therefore, market makers are generally experienced in 
dealing with the difference between their model deltas (which is their
submission to the dealer poll) and market deltas (which is the average
of the dealer poll).

\begin{table}
\caption{Model Delta Comparison\label{delta}}

\center
\footnotesize

\begin{tabular}{|rr|rr|rr|rr|rr|}
\hline
\multicolumn{ 2}{|c|}{{\bf CDX-IG9}} & \multicolumn{ 2}{|c|}{{\bf BC-AH}} & \multicolumn{ 2}{|c|}{{\bf BC-CN}} & \multicolumn{ 2}{|c|}{{\bf DIC $\alpha = 0.2$}} & \multicolumn{ 2}{|c|}{{\bf DIC $\alpha = 1$}} \\
\hline
{\bf Att} & {\bf Det} & {\bf 5Y} & {\bf 7Y} & {\bf 5Y} & {\bf 7Y} & {\bf 5Y} & {\bf 7Y} & {\bf 5Y} & {\bf 7Y} \\
\hline
0.0\% & 2.4\% &        4.20  &        1.78  &        4.21  &        1.75  &        5.10  &        2.17  &        4.92  &        1.98  \\
2.4\% & 6.5\% &        7.83  &        4.67  &        7.70  &        4.58  &        8.53  &        5.44  &        8.17  &        5.13  \\
6.5\% & 9.6\% &        4.58  &        4.28  &        4.54  &        4.20  &        5.96  &        5.56  &        6.75  &        6.02  \\
9.6\% & 14.8\% &        2.39  &        2.70  &        2.41  &        2.64  &        2.93  &        5.12  &        2.64  &        4.96  \\
14.8\% & 30.3\% &        0.68  &        1.11  &        0.68  &        1.13  &        1.01  &        1.07  &        1.09  &        1.18  \\
30.3\% & 61.2\% &        0.20  &        0.55  &        0.23  &        0.59  &        0.06  &        0.13  &        0.05  &        0.10  \\
\hline
\end{tabular}

\vspace{.5cm}

\begin{tabular}{|rr|rr|rr|rr|rr|}
\hline
\multicolumn{ 2}{|c|}{{\bf iTraxx-S9}} & \multicolumn{ 2}{|c|}{{\bf BC-AH}} & \multicolumn{ 2}{|c|}{{\bf BC-CN}} & \multicolumn{ 2}{|c|}{{\bf DIC $\alpha = 0.2$}} & \multicolumn{ 2}{|c|}{{\bf DIC $\alpha = 1$}} \\
\hline
{\bf Att} & {\bf Det} & {\bf 5Y} & {\bf 7Y} & {\bf 5Y} & {\bf 7Y} & {\bf 5Y} & {\bf 7Y} & {\bf 5Y} & {\bf 7Y} \\
\hline
0.0\% & 3.0\% &      11.37  &        5.98  &      11.18  &        5.84  &      16.26  &        8.59  &      16.20  &        8.47  \\
3.0\% & 6.0\% &        5.22  &        5.04  &        5.12  &        4.93  &        4.15  &        5.43  &        4.10  &        5.39  \\
6.0\% & 9.0\% &        3.37  &        3.70  &        3.30  &        3.71  &        4.12  &        5.27  &        4.19  &        5.22  \\
9.0\% & 12.0\% &        1.80  &        2.52  &        1.96  &        2.33  &        3.85  &        4.84  &        3.85  &        5.00  \\
12.0\% & 22.0\% &        0.92  &        1.22  &        0.92  &        1.17  &        0.84  &        1.12  &        0.86  &        1.27  \\
22.0\% & 60.0\% &        0.35  &        0.58  &        0.38  &        0.64  &        0.07  &        0.26  &        0.07  &        0.24  \\
\hline
\end{tabular}

\vspace{.5cm}

\begin{tabular}{|rr|rr|rr|rr|rr|}
\hline
\multicolumn{ 2}{|c|}{{\bf CDX-HY9}} & \multicolumn{ 2}{|c|}{{\bf BC-AH}} & \multicolumn{ 2}{|c|}{{\bf BC-CN}} & \multicolumn{ 2}{|c}{{\bf DIC $\alpha = 0.2$}} & \multicolumn{ 2}{|c|}{{\bf DIC $\alpha = 1$}} \\
\hline
{\bf Att} & {\bf Det} & {\bf 5Y} & {\bf 7Y} & {\bf 5Y} & {\bf 7Y} & {\bf 5Y} & {\bf 7Y} & {\bf 5Y} & {\bf 7Y} \\
\hline
0.0\% & 4.0\% &        2.03  &        0.62  &        1.96  &        0.60  &        1.92  &        0.53  &        1.96  &        0.60  \\
4.0\% & 15.6\% &        2.90  &        1.63  &        2.86  &        1.62  &        3.33  &        2.22  &        3.05  &        1.88  \\
15.6\% & 27.2\% &        1.74  &        1.49  &        1.74  &        1.48  &        2.11  &        1.37  &        2.40  &        1.63  \\
27.2\% & 56.3\% &        0.79  &        1.63  &        0.83  &        1.65  &        0.69  &        1.29  &        0.66  &        1.34  \\
\hline
\end{tabular}

\end{table}

Table \ref{delta} showed the additive model deltas from the 
calibrated DIC model computed by keeping the marginal distribution 
of $X_t$ constant, as well as the model deltas from 
the base correlation model. The deltas in table 
\ref{delta} is the ratio between the change in tranche ETL and the change 
in index swap expected loss. This ratio is different from the usual tranche 
delta definition as the ratio of PV changes. We choose to compare the 
models' expected loss deltas since they do not depend on the coupon 
payments of the tranche, thus making it easier to compare across 
capital structure. 

Different stochastic recovery models only make a very small difference 
on the model deltas under the base correlation model, so is different 
$\alpha$ values under the DIC model. The model deltas from the DIC model 
are roughly in line with the model deltas from the base correlation
except that the DIC deltas tend to be lower for senior tranches and 
higher for equity tranches for the two investment grade indices. 
Certain tweaks to the DIC model can bring its model deltas closer to the 
BC-TLP model deltas. For example, a ``armageddon'' event can be introduced, 
under which all the credits in a portfolio default together. If 
a portion of the index swap spread bump is attributed to the 
increase in the probability of the armageddon event, then the deltas 
in the senior tranches would become higher and equity tranche would 
become lower, which brings the DIC model deltas closer to the BC-TLP 
delta. The idea of armageddon event is not new, for example, it appeared 
in an early work of \cite{compbasket}. However, it is questionable 
whether the practical benefits from such a tweak, whose sole purpose 
is to bring the DIC model deltas closer to the BC-TLP model deltas, 
would outweigh the added complexity to the model, especially given 
that the market makers are already experienced in dealing with 
the difference between model deltas and market deltas. 

\subsection{Price Index Tranches as Bespoke}

\begin{table}
\caption{Price Index Tranches as Bespoke Tranches to CDX-IG9\label{idxasbspk}}
\center
\footnotesize

\begin{tabular}{|rr|rr|rr|rr|rr|rr|}
\hline
\multicolumn{ 2}{|c|}{\multirow{2}{*}{\bf iTraxx-S9}} & \multicolumn{ 2}{|c|}{\multirow{2}{*}{\bf Market ETL}} & \multicolumn{ 8}{|c|}{{\bf Difference from Market ETL if Priced as Bespoke}} \\
\cline{5-12}
\multicolumn{ 2}{|c}{{\bf }} & \multicolumn{ 2}{|c}{{\bf }} & \multicolumn{ 2}{|c|}{{\bf BC-AH}} & \multicolumn{ 2}{|c|}{{\bf BC-CN}} & \multicolumn{ 2}{|c|}{{\bf DIC $\alpha=0.2$}} & \multicolumn{ 2}{|c|}{{\bf DIC $\alpha=1$}} \\
\hline
{\bf Att} & {\bf Det} & {\bf 5Y} & {\bf 7Y} & {\bf 5Y} & {\bf 7Y} & {\bf 5Y} & {\bf 7Y} & {\bf 5Y} & {\bf 7Y} & {\bf 5Y} & {\bf 7Y} \\
\hline
0.0\% & 3.0\% & 35.59\% & 56.69\% & -1.08\% & -1.54\% & -0.79\% & -1.60\% & -0.31\% & -2.54\% & -0.82\% & -2.89\% \\
3.0\% & 6.0\% & 7.73\% & 22.06\% & 4.20\% & 4.05\% & 4.21\% & 4.31\% & 2.25\% & 2.18\% & 2.80\% & 3.06\% \\
6.0\% & 9.0\% & 4.56\% & 13.46\% & 0.63\% & 1.82\% & 0.46\% & 1.96\% & -0.21\% & -0.65\% & -0.32\% & -0.53\% \\
9.0\% & 12.0\% & 2.27\% & 7.31\% & 0.23\% & 1.85\% & 0.25\% & 1.72\% & 0.25\% & -0.78\% & 0.35\% & -0.57\% \\
12.0\% & 22.0\% & 0.80\% & 3.02\% & 0.24\% & 0.36\% & 0.21\% & 0.33\% & 0.25\% & -0.12\% & 0.27\% & 0.07\% \\
22.0\% & 60.0\% & 0.51\% & 1.84\% & -0.33\% & -0.64\% & -0.33\% & -0.63\% & -0.16\% & 0.05\% & -0.16\% & -0.03\% \\
\hline
\multicolumn{ 2}{|c|}{{\bf RMS}} &   0.00\%    &  0.00\%     & 1.80\% & 2.08\% & 1.77\% & 2.18\% & 0.94\% & 1.43\% & 1.21\% & 1.75\% \\
\hline
\end{tabular}

\vspace{.5cm}

\begin{tabular}{|rr|rr|rr|rr|rr|rr|}
\hline
\multicolumn{ 2}{|c|}{\multirow{2}{*}{\bf CDX-HY9}} & \multicolumn{ 2}{|c|}{\multirow{2}{*}{\bf Market ETL}} & \multicolumn{ 8}{|c|}{{\bf Difference from Market ETL if Priced as Bespoke}} \\
\cline{5-12}
\multicolumn{ 2}{|c}{{\bf }} & \multicolumn{ 2}{|c}{{\bf }} & \multicolumn{ 2}{|c|}{{\bf BC-AH}} & \multicolumn{ 2}{|c|}{{\bf BC-CN}} & \multicolumn{ 2}{|c|}{{\bf DIC $\alpha=0.2$}} & \multicolumn{ 2}{|c|}{{\bf DIC $\alpha=1$}} \\
\hline
{\bf Att} & {\bf Det} & {\bf 5Y} & {\bf 7Y} & {\bf 5Y} & {\bf 7Y} & {\bf 5Y} & {\bf 7Y} & {\bf 5Y} & {\bf 7Y} & {\bf 5Y} & {\bf 7Y} \\
\hline
0.0\% & 4.0\% & 78.29\% & 91.76\% & 0.44\% & 2.36\% & 0.70\% & 2.18\% & -1.75\% & -0.52\% & 0.93\% & 2.58\% \\
4.0\% & 15.6\% & 36.47\% & 62.78\% & 3.63\% & 5.63\% & 3.66\% & 5.47\% & -1.04\% & 0.24\% & -2.02\% & -1.17\% \\
15.6\% & 27.2\% & 13.20\% & 32.05\% & 0.07\% & 5.95\% & -0.05\% & 5.92\% & 2.72\% & 6.26\% & 3.76\% & 8.07\% \\
27.2\% & 56.3\% & 4.55\% & 21.63\% & -1.72\% & -11.08\% & -1.71\% & -10.98\% & 0.47\% & -2.64\% & 0.10\% & -3.17\% \\
\hline
\multicolumn{ 2}{|c|}{{\bf RMS}} &  0.00\%     & 0.00\%      & 2.02\% & 6.99\% & 2.05\% & 6.90\% & 1.71\% & 3.41\% & 2.18\% & 4.56\% \\
\hline
\end{tabular}

\end{table}

Standard tranches on one index can also be viewed as bespoke 
tranches to another index, it is interesting to compare the observed 
index tranche prices against the model prices if they were valued as 
bespoke tranches to another index. It is certainly preferable if the market 
prices are close to the ``as bespoke'' model prices since it means 
that the model's  response to changes in the underlying credit portfolio 
is similar to the market's response. 

In this study, we priced the CDX-HY9 and iTraxx-S9 as bespoke tranches  
to the CDX-IG9 using both the DIC model and the standard BC-TLP model. 
CDX-IG9 is chosen to be the ``index'' because it has similar properties to 
both iTraxx-S9 and CDX-HY9: the CDX-IG9 and iTraxx-S9 are both investment 
grade indices and the CDX-IG9 and CDX-HY9 are both north American names. 
In comparison, the iTraxx-S9 and CDX-HY9 indices are more dissimilar to
each other.
 
Table \ref{idxasbspk} showed the difference
between the market ETLs and the ETLs priced as bespoke 
tranches to the CDX-IG9. Interestingly, the differences between the
market and ``as bespoke'' ETL from the DIC model are generally smaller 
than those from the BC-TLP model. The performance of DIC model is quite 
promising in this study considering that the BC-TLP method has been 
long established as the market standard.

\subsection{Bespoke Tranche Pricing}
In this example, we show the bespoke pricing results for a super mixed 
portfolio, which is a union of the CDX-IG9, iTraxx-S9 and CDX-HY9
portfolios. Each standard index portfolio makes up 1/3 of the total notional 
in the super mixed portfolio. Since all the names in this super mix
portfolio appear in standard indices, the many-to-one restriction is
satisfied therefore we can use the fast simulation method described in
section \ref{many1}. Table \ref{supermix} showed the prices from 
the DIC-SAMC and the BC-TLP. The DIC-SAMC method uses $\alpha = 0.2$, 
and the BC-TLP method uses the stochastic recovery model from \cite{hitier}. 

\begin{table}
\caption{SuperMix Bespoke Portfolio\label{supermix}}
\center

\underline{5Y Bespoke ETL}

\vspace{.25cm}

\footnotesize
\begin{tabular}{|cc|rrrrrr|rrr|rr|}
\hline
\multicolumn{ 2}{|c|}{{\bf SuperMix}} & \multicolumn{ 6}{|c|}{{\bf DIC-SAMC: Market Factor Correlations}} & \multicolumn{ 5}{|c|}{{\bf BC-TLP with Different Mapping}} \\
\hline
{\bf Att} & {\bf Det} & {\bf 0} & {\bf 0.2} & {\bf 0.4} & {\bf 0.6} & {\bf 0.8} & {\bf 1} & {\bf CDX} & {\bf ITX} & {\bf HY9} & {\bf N\_WA} & {\bf R\_WA} \\
\hline
0\%   & 3\%   & 82.97\% & 80.50\% & 78.16\% & 76.00\% & 74.07\% & 72.25\% & 71.04\% & 75.48\% & 69.88\% & 72.13\% & 70.78\% \\
3\%   & 7\%   & 36.68\% & 36.12\% & 35.32\% & 34.37\% & 33.46\% & 32.53\% & 35.76\% & 29.93\% & 32.05\% & 32.58\% & 32.56\% \\
7\%   & 10\%  & 15.42\% & 16.09\% & 16.44\% & 16.54\% & 16.47\% & 16.12\% & 18.88\% & 14.22\% & 16.58\% & 16.56\% & 16.77\% \\
10\%  & 15\%  & 7.49\% & 8.00\% & 8.40\% & 8.71\% & 8.94\% & 8.88\% & 8.80\% & 7.95\% & 9.26\% & 8.67\% & 9.01\% \\
15\%  & 30\%  & 2.92\% & 3.14\% & 3.31\% & 3.45\% & 3.53\% & 3.69\% & 2.52\% & 2.39\% & 3.73\% & 2.88\% & 3.33\% \\
30\%  & 60\%  & 0.05\% & 0.13\% & 0.26\% & 0.46\% & 0.65\% & 0.77\% & 0.52\% & 1.56\% & 0.93\% & 1.00\% & 0.92\% \\
\hline
\end{tabular}

\vspace{.5cm}

\normalsize
\underline{7Y Bespoke ETL}

\vspace{.25cm}

\footnotesize
\begin{tabular}{|cc|rrrrrr|rrr|rr|}
\hline
\multicolumn{ 2}{|c|}{{\bf SuperMix}} & \multicolumn{ 6}{|c|}{{\bf DIC-SAMC: Market Factor Correlations}} & \multicolumn{ 5}{|c|}{{\bf BC-TLP with Different Mapping}} \\
\hline
{\bf Att} & {\bf Det} & {\bf 0} & {\bf 0.2} & {\bf 0.4} & {\bf 0.6} & {\bf 0.8} & {\bf 1} & {\bf CDX} & {\bf ITX} & {\bf HY9} & {\bf N\_WA} & {\bf R\_WA} \\
\hline
0\%   & 3\%   & 96.22\% & 94.49\% & 92.80\% & 91.13\% & 89.52\% & 88.05\% & 89.32\% & 91.60\% & 86.14\% & 90.26\% & 87.46\% \\
3\%   & 7\%   & 70.88\% & 67.94\% & 65.25\% & 62.86\% & 60.76\% & 58.81\% & 62.97\% & 61.03\% & 57.47\% & 60.90\% & 58.99\% \\
7\%   & 10\%  & 42.93\% & 41.97\% & 40.73\% & 39.36\% & 37.92\% & 35.63\% & 43.06\% & 38.07\% & 36.27\% & 37.68\% & 37.81\% \\
10\%  & 15\%  & 30.99\% & 30.62\% & 29.95\% & 29.16\% & 28.50\% & 27.43\% & 27.85\% & 24.26\% & 23.26\% & 26.95\% & 24.27\% \\
15\%  & 30\%  & 12.99\% & 13.57\% & 13.92\% & 14.14\% & 14.62\% & 15.37\% & 8.61\% & 8.66\% & 16.63\% & 12.72\% & 14.05\% \\
30\%  & 60\%  & 0.58\% & 1.04\% & 1.57\% & 2.07\% & 2.45\% & 2.31\% & 3.79\% & 4.95\% & 3.52\% & 3.11\% & 3.76\% \\
\hline
\end{tabular}

\end{table}

In the DIC-SAMC method, we assumed the pair-wise correlation between the three 
market factors are the same even though it is very easy to accommodate a full 
correlation matrix. It is very intuitive in table 
\ref{supermix} that the DIC-SAMC produces lower equity tranche ETLs and 
higher senior tranche ETLs with higher correlations between market factors. 
In the BC-TLP section of table \ref{supermix}, we reported the resulting 
ETLs from mappings to
each individual index, as well as the notional and risk weighted average ETL 
in the ``N\_WA'' and ``R\_WA'' columns. The notional and risk weighted average 
of the BC-TLP ETLs are very close to the DIC-SAMC results with very high
correlation between 80\% and 100\%. This is somewhat expected because the BC-TLP 
is still a one factor model even though it is mapped to three indices
separately. The effects of lower correlations between market factors 
cannot be produced from such a one-factor model. On the other hand, the 
DIC-SAMC method is able to produce the pricing effects of lower correlations 
between market factors. The one-factor limitation of BC-TLP is not yet a problem 
in practice since market participants do expect different market factors to be 
highly correlated. For example, table \ref{idxcorr} showed the correlations 
between the index swap spread returns estimated from historical weekly index 
spreads from Mar 2008 to Feb 2010. The correlations are all in the mid 90\% 
between the three main indices. However, being able to price with low market 
factor correlation is certainly an important advantage in DIC-SAMC since it 
helps the market participants to gauge the correlation risk between market 
factors, and allows them to manage or trade such risk. 

\begin{table}
\caption{Correlation between Index Swap Spread Returns\label{idxcorr}}
\center
\footnotesize

\begin{tabular}{|r|rrr|}
\hline
      & {\bf CDX-IG9} & {\bf CDX-HY9} & {\bf Itraxx-S9} \\
\hline
{\bf CDX-IG9} & 1     & 0.9328 & 0.9625 \\
{\bf CDX-HY9} & 0.9328 & 1     & 0.9531 \\
{\bf Itraxx-S9} & 0.9625 & 0.9531 & 1 \\
\hline
\end{tabular}

\end{table}

Since the normal approximation in the DIC-SAMC could result in a pricing 
error of up to 0.1\%\footnote{see \cite{self} for more details.}, there is 
no benefit to run the DIC-SAMC method with higher accuracy than 0.1\%. It only takes 
about 250K simulation paths in DIC-SAMC for the pricing error to be less than 0.1\% 
for all tranches even without using any variance reduction technique. The left 
half in table \ref{mcerr} showed the simulation error with 250K path and no 
variance reduction, in which case the simulation error is almost the same 
for different correlation values among market factors. With 250K path, the 
DIC-SAMC takes less than a second to compute the ETL at a given time horizon 
on a regular PC; thus a typical bespoke tranche pricing only takes a few seconds 
with DIC-SAMC assuming that the simulation needs to be run on every quarterly 
date.

\begin{table}
\caption{DIC-SAMC Error from 250K Simulation Paths\label{mcerr}}
\center

\vspace{.25cm}

\footnotesize

\begin{tabular}{|cc|cc|cc|cc|cc|}
\hline
\multicolumn{ 2}{|c|}{\multirow{2}{*}{{\bf SuperMix}}} & \multicolumn{ 4}{|c|}{{\bf Regular DIC-SAMC}} & \multicolumn{ 4}{|c|}{{\bf DIC-SAMC w/ Control Variate}} \\
\cline{3-10}
\multicolumn{ 2}{|c|}{{\bf }} & \multicolumn{ 2}{|c|}{{\bf Absolute Err}} & \multicolumn{ 2}{|c|}{{\bf Relative Err}} & \multicolumn{ 2}{|c|}{{\bf Absolute Err}} & \multicolumn{ 2}{|c|}{{\bf Relative Err}} \\
\hline
{\bf Att} & {\bf Det} & {\bf 5Y} & {\bf 7Y} & {\bf 5Y} & {\bf 7Y} & {\bf 5Y} & {\bf 7Y} & {\bf 5Y} & {\bf 7Y} \\
\hline
0.00\% & 3.00\% & 0.056\% & 0.036\% & 0.077\% & 0.041\% & 0.019\% & 0.016\% & 0.026\% & 0.018\% \\
3.00\% & 7.00\% & 0.080\% & 0.078\% & 0.245\% & 0.133\% & 0.027\% & 0.026\% & 0.083\% & 0.044\% \\
7.00\% & 10.00\% & 0.067\% & 0.089\% & 0.413\% & 0.251\% & 0.028\% & 0.036\% & 0.175\% & 0.100\% \\
10.00\% & 15.00\% & 0.050\% & 0.088\% & 0.567\% & 0.320\% & 0.023\% & 0.046\% & 0.256\% & 0.169\% \\
15.00\% & 30.00\% & 0.032\% & 0.056\% & 0.858\% & 0.362\% & 0.016\% & 0.021\% & 0.442\% & 0.139\% \\
30.00\% & 60.00\% & 0.009\% & 0.020\% & 1.168\% & 0.875\% & 0.007\% & 0.014\% & 0.973\% & 0.606\% \\
\hline
\end{tabular}

\end{table}

A number of variance reduction techniques could be applied to further improve the computational 
speed, for example the quasi-random sequence and importance sampling could be effective. In particular,
the 100\% correlation case can be used as an effective control variate for the DIC-SAMC. With 
100\% correlation, the multiple market factors reduce to a single market factor, therefore 
bespoke tranches can be priced using a regular one-factor semi-analytical method without 
simulation. This control variate is particularly effective if the correlations between market 
factors are high, which is indeed the case in practice. The right half in Table \ref{mcerr} 
showed the simulation error with 90\% correlation between market factors using the control 
variate. The simulation error has been significantly reduced by the control variate, which 
translates to roughly 3-5 times reduction in computation time. 

\subsection{Single Name Deltas}
The single name deltas are the most time consuming risk measures to compute for bespoke tranches.
Using the DIC-SAMC method, the path-wise single name deltas can be computed analytically from 
the normal approximation, therefore, it is very fast to compute the single name deltas by 
integrating the path-wise deltas directly within the DIC-SAMC simulation. In the SuperMix 
portfolio of the previous example, the path-wise delta integration method only takes about 
5 seconds to achieve reasonable convergence for all the single name deltas at a given tenor. 
Overall, the PV and single name risks for a single tranche only take several minutes to 
compute on a regular PC, therefore the DIC-SAMC method is fast enough to support a large 
trading book in practice with a small server farm. For example, a trading book of 2,000 
bespoke trades would take about 4 hours to compute PVs and single name risks using a
server farm with 40 CPUs.

\begin{figure}
\caption{Single Name Hedge Ratios for the SuperMix Portfolio \label{sndelta}}

\vspace{.25cm}
\center
\begin{minipage}{3in}
\center
\underline{DIC-SAMC 5Y 0-3\%}
\scalebox{.55}{\includegraphics{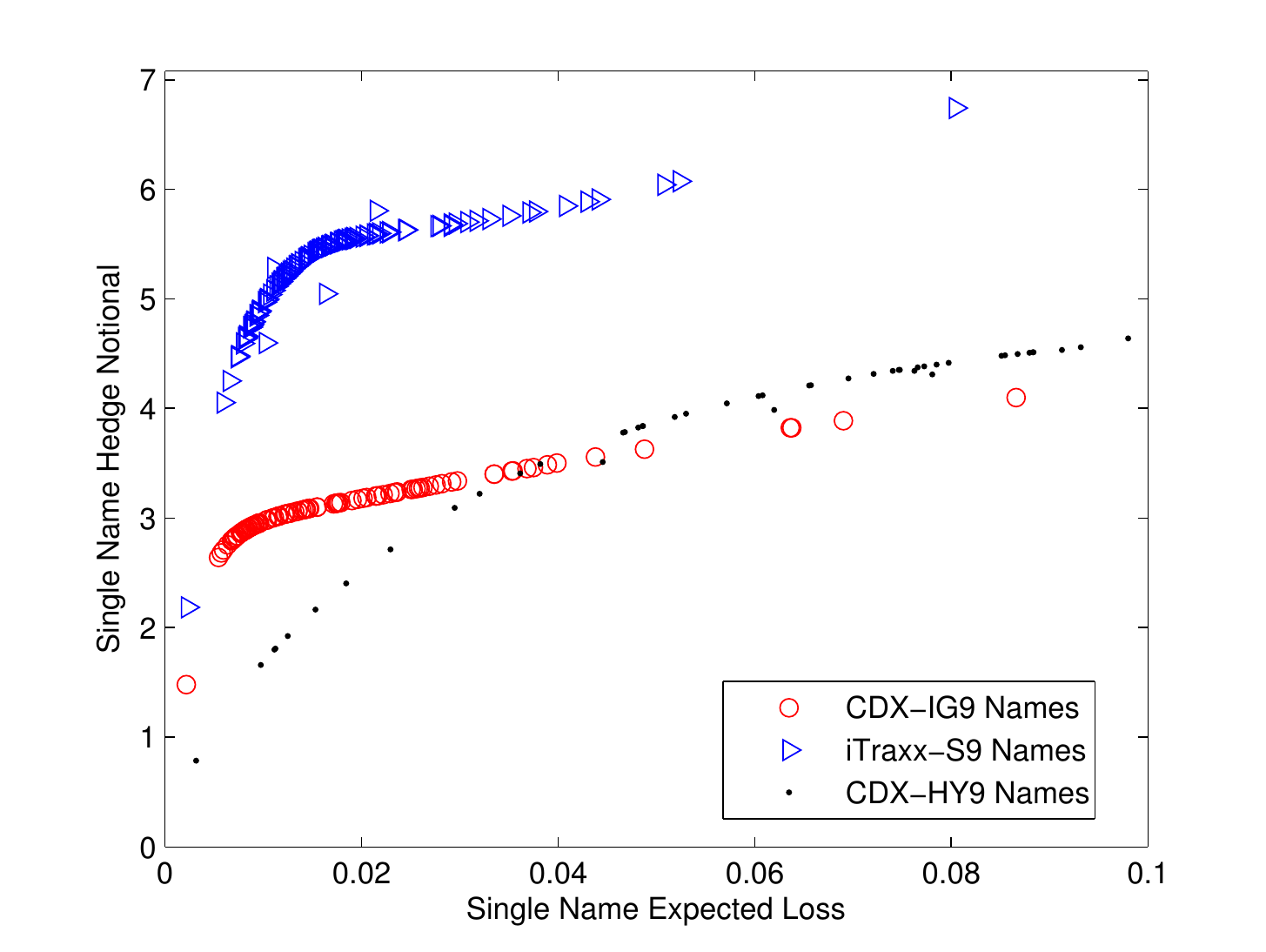}}
\end{minipage}
\begin{minipage}{3in}
\center
\underline{BC-TLP 5Y 0-3\%}
\scalebox{.55}{\includegraphics{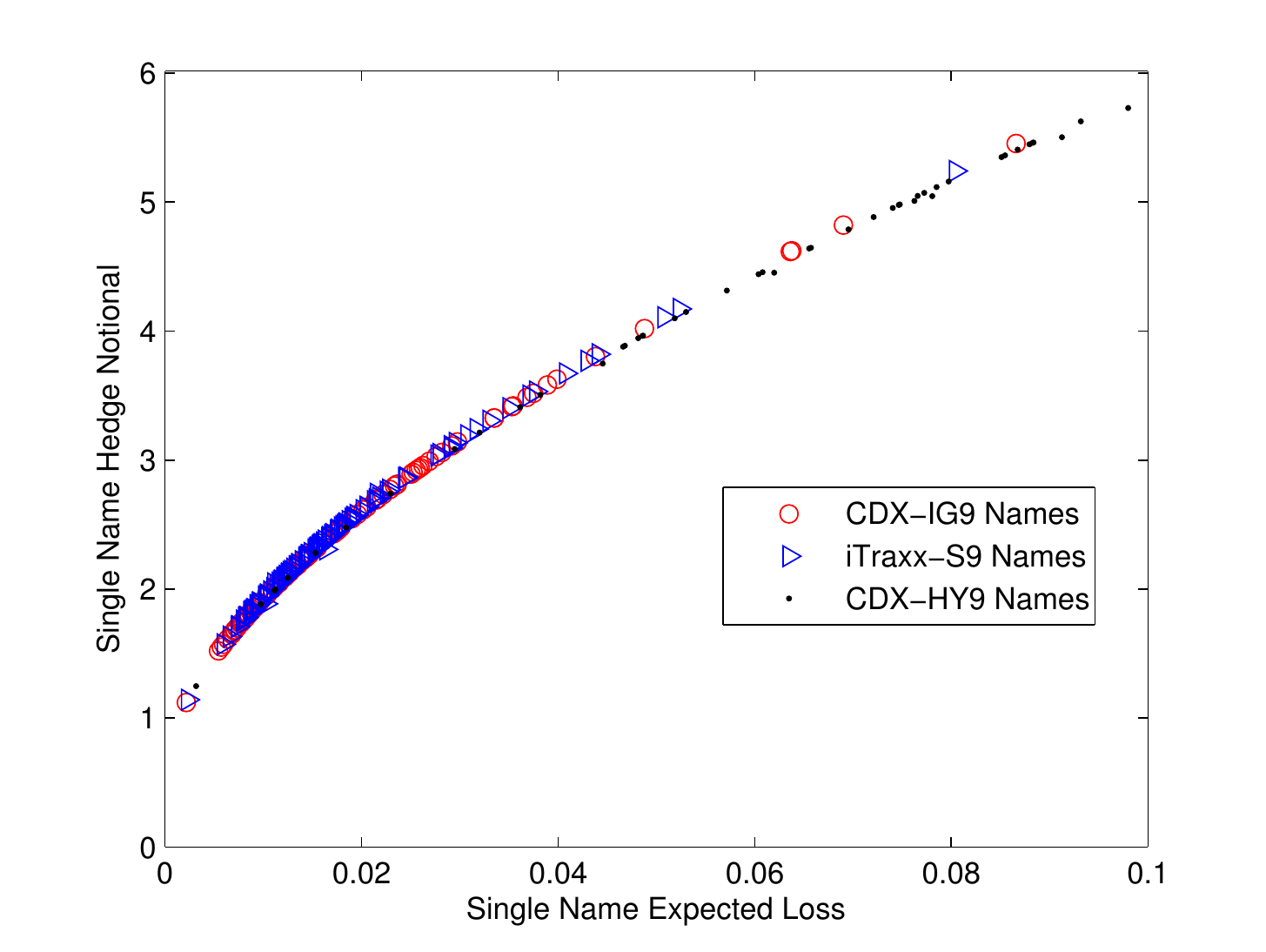}}
\end{minipage}

\begin{minipage}{3in}
\center
\underline{DIC-SAMC 5Y 7-10\%}
\scalebox{.55}{\includegraphics{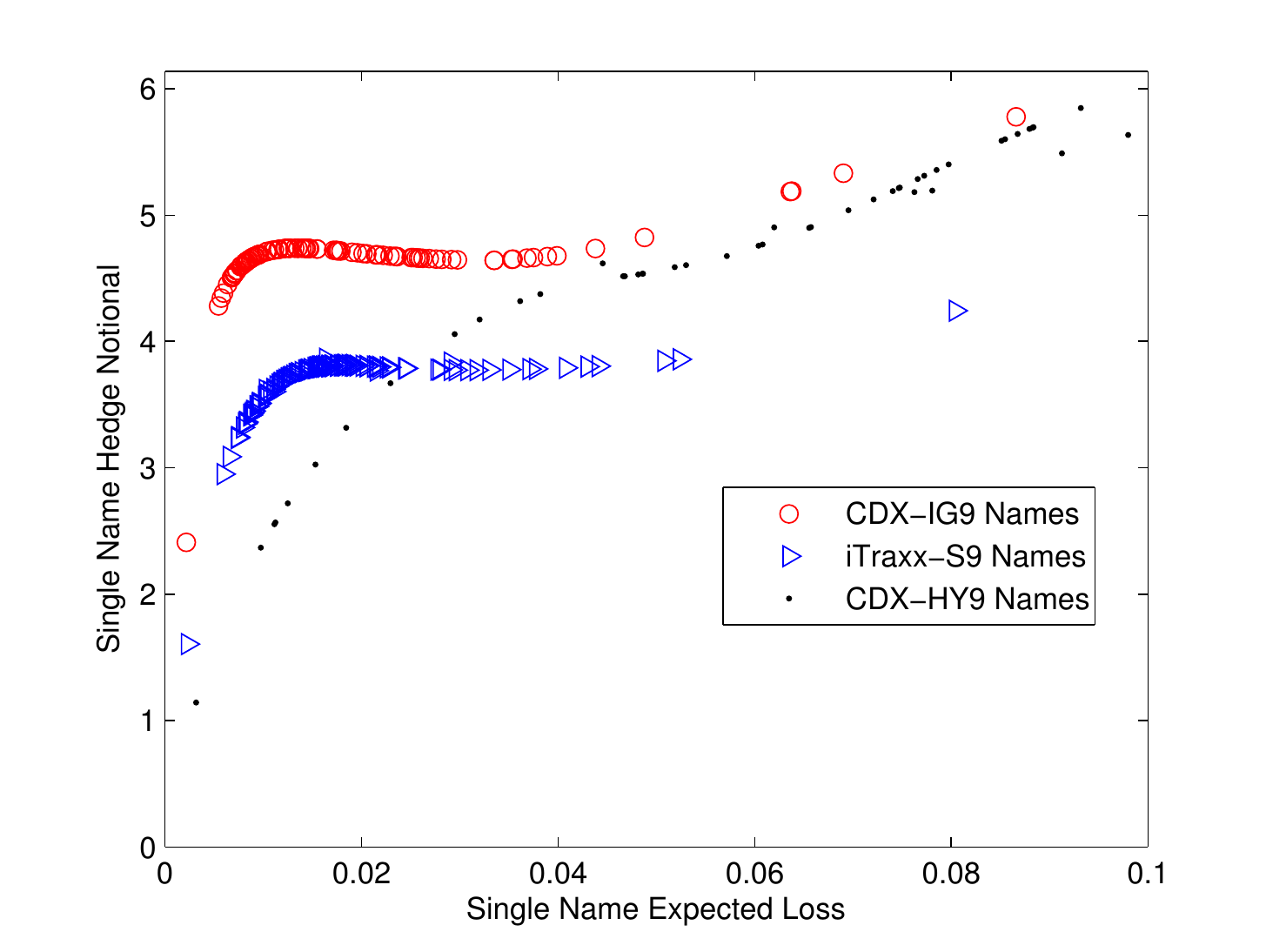}}
\end{minipage}
\begin{minipage}{3in}
\center
\underline{BC-TLP 5Y 7-10\%}
\scalebox{.55}{\includegraphics{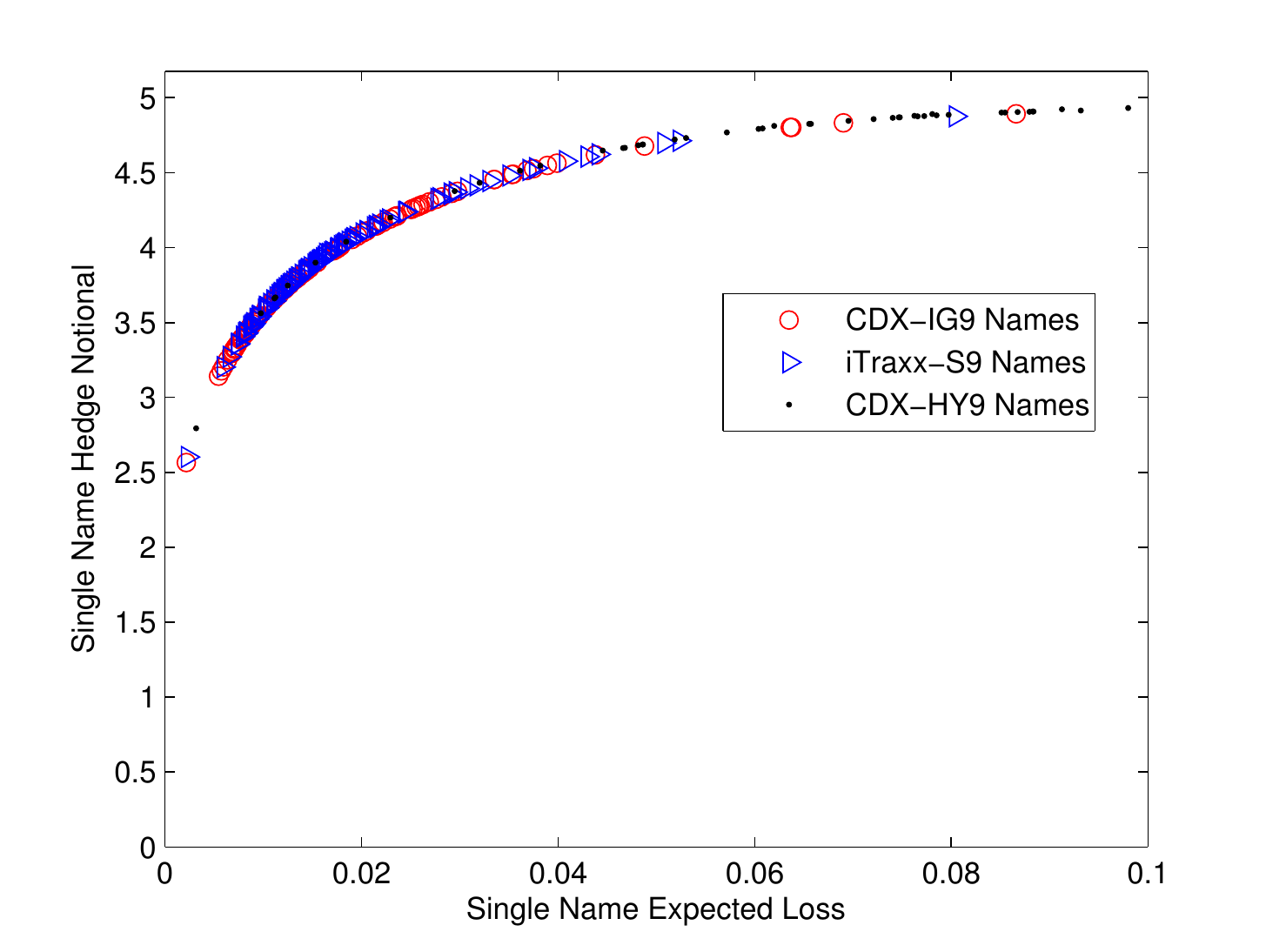}}
\end{minipage}

\begin{minipage}{3in}
\center
\underline{DIC-SAMC 5Y 30-60\%}
\scalebox{.55}{\includegraphics{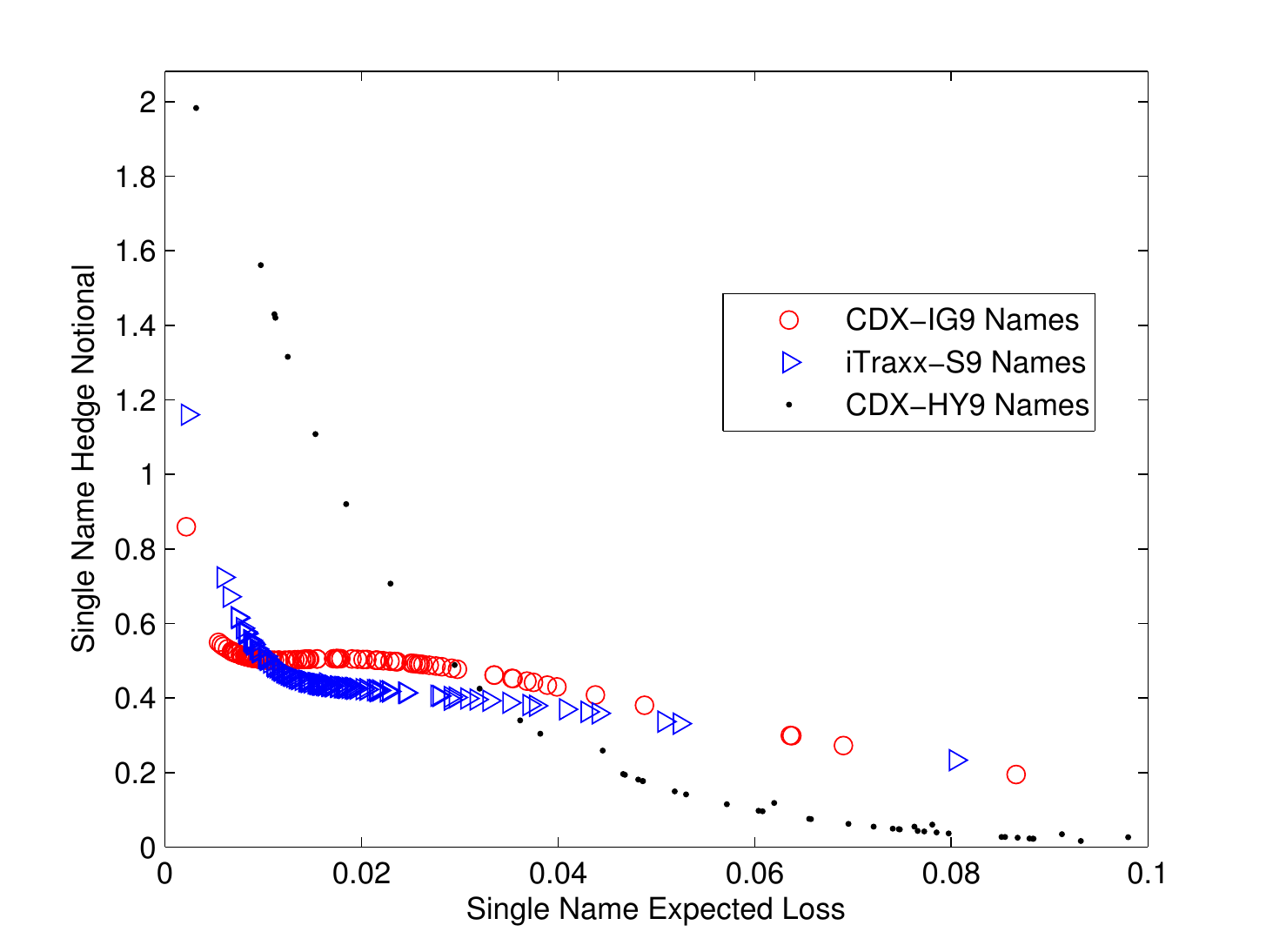}}
\end{minipage}
\begin{minipage}{3in}
\center
\underline{BC-TLP 5Y 30-60\%}
\scalebox{.55}{\includegraphics{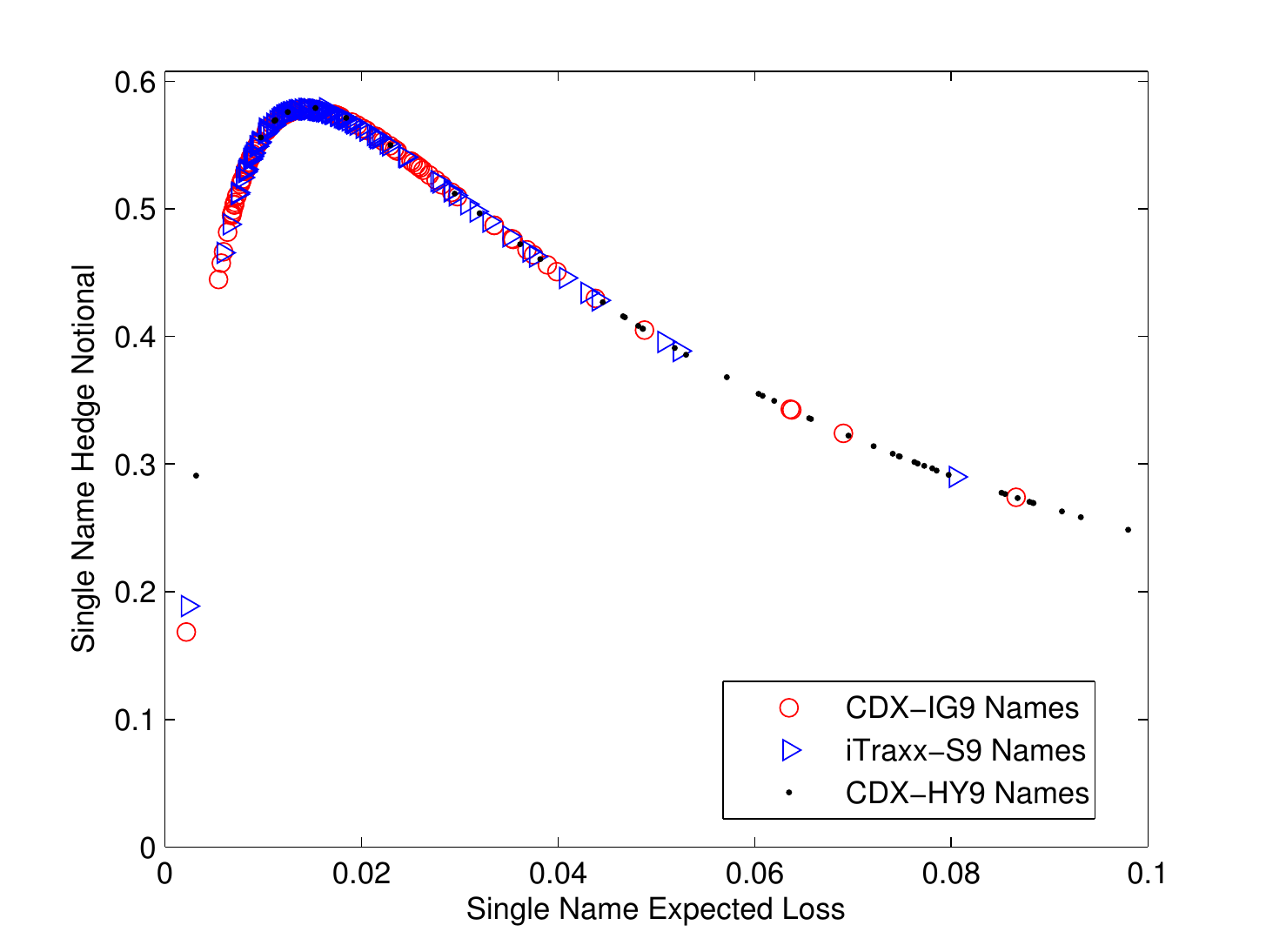}}
\end{minipage}

\end{figure}

Figure \ref{sndelta} showed the single name hedge ratios from both the DIC-SAMC and BC-TLP mothods
for a few tranches on the SuperMix portfolio with 90\% correlation between market factors. The 
hedge ratios are computed as the ratio of change 
in tranche expected loss over the single name expected loss for a small perturbation in the
single name default probability. Since the single name notionals are not uniform in the SuperMix 
portfolio, we normalized the resulting hedge ratios by each single name's notional amount in the 
SuperMix portfolio in order to compare the risk contribution from a unit amount of single name 
notional. A few interesting observations can be made from figure \ref{sndelta}:
\begin{compactitem}
\item The single name risks from the DIC-SAMC methods depend on the name's associated market factor. 
Names from different credit indices can have different single name risks even if their default probability 
and recovery rates are identical. This is a natural consequence of the multiple market 
factors and the identity consistency in the DIC-SAMC model. The BC-TLP single name deltas do not 
depends on the credit index association as BC-TLP is a one-factor model without identity
consistency.
\item In DIC-SAMC, the junior tranches are more sensitive to riskier names and senior tranches 
are more sensitive to safer names, which is quite intuitive. In BC-TLP, the 30-60\% 
tranche is much more sensitive to names with expected loss between 1-2\%, which is 
rather counter-intuitive and difficult to explain. 
\item The deltas in DIC-SAMC are always positive by construction (subject to the Monte Carlo 
simulation error), while 
the BC-TLP could produce negative deltas. Even though there is no negative single name deltas 
shown in table \ref{sndelta}, the BC-TLP model did produce some negative deltas for the SuperMix
portfolio when mapped to the iTraxx-S9 index.
\end{compactitem}

\begin{table}
\caption{Fixed Recovery Tranche \label{fixrec}}
\center

\normalsize
\underline{DIC-SAMC 5Y}

\vspace{.25cm}
\footnotesize
\begin{tabular}{|cc|rrr|rrr|}
\hline
\multicolumn{ 2}{|c}{{\bf SuperMix}} & \multicolumn{ 3}{|c|}{{\bf Mkt Factor Corr = 0.8}} & \multicolumn{ 3}{|c|}{{\bf Mkt Factor Corr = 1}} \\
\hline
       Att &        Det &  Rec = 0.2 &  Rec = 0.4 &  Rec = 0.6 &  Rec = 0.2 &  Rec = 0.4 &  Rec = 0.6 \\
\hline
       0\% &        3\% &    80.50\% &    74.47\% &    63.84\% &    78.70\% &    72.68\% &    62.20\% \\

       3\% &        7\% &    43.39\% &    32.78\% &    19.51\% &    42.27\% &    31.89\% &    19.05\% \\

       7\% &       10\% &    23.02\% &    15.08\% &     7.15\% &    22.66\% &    14.73\% &     7.32\% \\

      10\% &       15\% &    12.77\% &     7.55\% &     3.50\% &    12.64\% &     7.62\% &     3.76\% \\

      15\% &       30\% &     4.57\% &     2.47\% &     0.83\% &     4.86\% &     2.78\% &     0.98\% \\

      30\% &       60\% &     0.83\% &     0.33\% &     0.04\% &     1.03\% &     0.43\% &     0.11\% \\
\hline
\end{tabular}  

\vspace{.5cm}

\normalsize
\underline{BC-TLP 5Y}
\vspace{.25cm}
\footnotesize

\begin{tabular}{|cc|rrr|rrr|}
\hline
\multicolumn{ 2}{|c}{{\bf SuperMix}} & \multicolumn{ 3}{|c|}{{\bf Map with Stoch Recovery}} & \multicolumn{ 3}{|c|}{{\bf Map with Fixed Recovery}} \\
\hline
       Att &        Det &  Rec = 0.2 &  Rec = 0.4 &  Rec = 0.6 &  Rec = 0.2 &  Rec = 0.4 &  Rec = 0.6 \\
\hline
       0\% &        3\% &    80.57\% &    76.50\% &    65.03\% &    80.29\% &    74.78\% &    65.22\% \\

       3\% &        7\% &    42.44\% &    32.25\% &    20.45\% &    40.12\% &    32.65\% &    21.36\% \\

       7\% &       10\% &    22.82\% &    14.65\% &     7.28\% &    17.21\% &    15.44\% &     7.62\% \\

      10\% &       15\% &    12.31\% &     7.06\% &     2.83\% &     6.84\% &     7.54\% &     2.75\% \\

      15\% &       30\% &     4.07\% &     1.98\% &     0.58\% &     7.94\% &     2.27\% &     0.31\% \\

      30\% &       60\% &     1.20\% &     0.54\% &     0.03\% &     0.70\% &     0.35\% &     0.00\% \\
\hline
\end{tabular}

\end{table}

\subsection{Fixed Recovery Tranches}
Table \ref{fixrec} showed the 5Y ETLs of the SuperMix portfolio priced with fixed 
recovery rates of 20\%, 40\% and 60\% using both the DIC-SAMC and BC-TLP. 
The BC-TLP uses the \cite{hitier} stochastic recovery model and notional
weighted average for mappings to different indices. Table \ref{fixrec} showed the 
DIC-SAMC prices with two different market factor correlations and BC-TLP prices 
using both mapping methods described in section \ref{bcm}. The first 
method uses stochastic market recovery rate in the TLP mapping, 
then prices the fixed recovery tranche with the same base correlation surface as the 
regular market recovery tranche; the second method uses the fixed recovery rate 
directly for the bespoke tranche in the TLP mapping. The DIC-SAMC prices of the 
fixed recovery trades are generally closer to the first BC-TLP method, which is not 
surprising because the first BC-TLP method preserves the identity consistency 
between the market and fixed recovery tranches.  Also notable is that the second 
BC-TLP mapping method produces an obvious arbitrage between the 10-15\% and 15-30\% 
tranches for 20\% fixed recovery.

\subsection{Quanto Adjustment}
The consistent DIC model can be useful in a number of practical situations
where BC-TLP meets its limitations. In this example, we show how to compute the 
quanto adjustment for index tranches denominated in a different currency from 
the index currency. The key factor in the quanto adjustment is the correlation 
between defaults and the exchange rates, which is difficult to capture 
properly under the base correlation framework. 

We have been ignoring the quanto adjustment for iTraxx names in the previous 
examples of the SuperMix portfolio. In this example, we'll show that the
quanto adjustment can be easily incorporated into the DIC-SAMC method by
adding the exchange rate as another market factor.

We use CDX-IG9 tranches whose protection notional is denominated
in EUR as an example. These quanto tranches are easier to price in
the US risk neutral measure since the underlying single name CDS curves in 
the CDX portfolio are constructed in this measure. The USD 
is expected to appreciate against EUR if the US economy 
is doing better than the European economy, and vise versa. Since 
what's important is the relative health of the two economy, it is a 
rather natural idea to correlate the USD/EUR exchange rate with the 
difference between the CDX market factor and the iTraxx market factor. 
However, since market factor $X_t^i$ in \eqref{condp} does not have an 
absolute scale as the DIC remains the same if we scale $X_t^i$ up and scale 
$\beta_j(t)$ down by the same factor, we instead choose to correlate 
the USD/EUR exchange rate with the difference between the conditional 
expected losses in the CDX and iTraxx portfolio with unit notional
amounts, which is a metric with an absolute scale. The difference
between conditional expected portfolio loss (DCEPL) can be written as:
\[
\textup{DCEPL}(\vec{X}) = \mathbb{E}[L^\textup{CDX} - L^\textup{iTraxx}|\vec{X}]
\]

The DCEPL appears rather peculiar since the difference 
between the portfolio loss (i.e., $\textup{DPL} = L^\textup{CDX} - L^\textup{iTraxx}$) 
seems to be a more natural quantity to correlate with the exchange rate. 
However, the DPL depends on idiosyncratic dynamics, which requires
the full default time simulation. The DCEPL, on the other hand, 
does not depends on the idiosyncratic dynamics, thus it can be 
easily incorporated into the DIC-SAMC procedure. Therefore, it is
much more efficient numerically to use DCEPL instead of DPL. 

The calculation for quanto adjustment using DCEPL thus involves two steps: 
the first step runs a DIC-SAMC simulation with both iTraxx and CDX
portfolios to construct the distribution of the DCEPL; 
the 2nd step runs the DIC-SAMC simulation again and uses a Gaussian 
Copula to generate the correlated USD/EUR exchange rate based on 
the DCEPL for each draw of $\vec{X}$; the exchange rate is then used to 
convert the ETL from EUR to USD. The distribution of the DCEPL from 
the first step is used as marginal distribution in the Gaussian 
Copula to draw the correlated USD/EUR exchange rate in the second
step.

Table \ref{quanto} showed the quanto adjustment amounts for CDX-IG9 5Y 
tranches assuming the USD/EUR exchange rate is lognormal with a cumulative 
volatility of 30\% at the 5Y maturity. The general features of the quanto 
adjustment from the DIC-SAMC simulation are quite intuitive: the quanto 
adjustments are generally positive and is more significant for senior tranches 
in relative scale because when CDX-IG9 suffers large losses, it is more likely 
that USD would depreciate, thus leading to more value in the protection 
denominated in EUR. Table \ref{quanto} also showed that the quanto 
adjustment is bigger with larger correlation between the EUR/USD exchange 
rate and the DCEPL. 

\begin{table}
\caption{Quanto Adjustment for CDX-IG9 Tranches\label{quanto}}
\center
\footnotesize

\begin{tabular}{|cc|c|rrrrr|}
\hline
\multicolumn{ 2}{|c|}{{\bf CDX-IG9}} & {\bf USD} & \multicolumn{5}{|c|}{{\bf Quanto Adj by Corr(FX, DCEPL)}} \\
\cline{1-2} \cline{4-8}
{\bf Att} & {\bf Det} & {\bf ETL} & {\bf 0.0} & {\bf 0.2} & {\bf 0.4} & {\bf 0.6} & {\bf 0.8} \\
\hline
0\%   & 2.4\%   & 67.23\% & 0.00\% & 1.26\% & 2.39\% & 3.85\% & 4.81\% \\
2.4\%   & 6.5\%   & 23.24\% & 0.00\% & 0.97\% & 1.84\% & 3.00\% & 3.92\% \\
6.5\%   & 9.6\%  & 7.53\% & 0.00\% & 0.45\% & 0.82\% & 1.36\% & 1.83\% \\
9.6\%  & 14.8\%  & 3.18\% & 0.00\% & 0.29\% & 0.48\% & 0.81\% & 1.04\% \\
14.8\%  & 30.3\%  & 0.83\% & 0.00\% & 0.10\% & 0.18\% & 0.30\% & 0.33\% \\
30.3\%  & 61.2\%  & 0.33\% & 0.00\% & 0.05\% & 0.11\% & 0.15\% & 0.20\% \\
\hline
0\%   & 100\% & 3.23\% & 0.00\% & 0.15\% & 0.27\% & 0.43\% & 0.55\% \\
\hline
\end{tabular}
\end{table}

\section{Conclusion}

This DIC-SAMC method proposed in this paper is fully consistent 
and arbitrage free. It preserves the single name's identity
consistency across all the index and bespoke tranches. All the 
modelling flaws in the current standard BC-TLP mapping method are 
addressed in the DIC-SAMC approach. The single name risks from 
DIC-SAMC are arguably better than those from the BC-TLP because 
of the absence of arbitrage and the explicit modelling of multiple 
market factors. The DIC-SAMC method is also computationally 
efficient, making it a viable alternative to the standard 
BC-TLP method. 

In the DIC-SAMC model setup, we assumed that there is a distinct
market factor for each liquid index. This one-to-one mapping 
between market factors and liquid indices greatly improves the 
numerical efficiency of the model calibration because each liquid
index can be calibrated separately. It is also easier to 
estimate the correlation between market factors from historical market data because
of this one-to-one association. The downside of this choice is 
that the market factors are not granular enough to capture
industry and sector information, it may not be suitable to 
price bespoke tranches that are concentrated in certain industries 
or sectors. However, bespoke portfolios are usually well diversified 
because they are constructed to achieve better ratings. The rating 
agency's rating criteria severely penalize non-diversified portfolios, 
therefore we rarely see any bespoke tranche with very 
concentrated portfolios in practice. 

The DIC-SAMC has less pricing uncertainty when comparing to
the very ad hoc BC-TLP. The bespoke tranche pricing uncertainties
under DIC-SAMC are mainly from two factors: the first is the 
correlation between market factors, the second is the factor loadings 
between bespoke names and index tranches. The meaning and effects of
these two factors are very intuitive and they can be easily estimated 
from historical time series of market observations. The BC-TLP method, 
on the other hand, has much more uncertainty in pricing because it 
depends on many ad hoc model parameters and assumptions, such as 
the interpolation and extrapolation methods on the base correlation curve, 
the weighting methods across indices etc. The TLP mapping itself is also 
questionable as alternative mapping methods such as ATM or PM could 
also be feasible in practice. Most of the parameters and model assumptions 
in the BC-TLP method do not have intuitive meaning; and their effects on 
the model price can be quite difficult to understand; and we can't
directly estimate them from historical time series of market 
observations. 

The DIC-SAMC method has the potential to bring much more pricing
transparency to bespoke CDO tranches, and hopefully the added pricing 
transparency will encourage more active trading in the bespoke 
tranche market.

\appendix

\bibliographystyle{apsr}
\bibliography{../creditref}

\end{document}